\documentclass[
aps,
prl,
twocolumn,
preprintnumbers,
longbibliography,
superscriptaddress,
amsmath,
amssymb,
floatfix]{revtex4-2}

\usepackage[%
  colorlinks=true,
  urlcolor=blue,
  linkcolor=magenta,
  citecolor=blue
]{hyperref}
\usepackage{graphicx} 
\usepackage{amsmath, amssymb, braket, dsfont}
\usepackage{xcolor, comment, url, xfrac}
\usepackage{xr-hyper} 
\usepackage{hyperref} 

\definecolor{dkblue}{rgb}{0,0,0.5}
\definecolor{dkred}{rgb}{0.5,0,0}

\date{\today}
\begin{document}
\title{Robustness of Kardar-Parisi-Zhang-like transport in long-range interacting \\ quantum spin chains}

\author{Sajant Anand}
\affiliation{Dept. of Physics, Harvard University, Cambridge, MA 02138, USA}
\affiliation{Dept. of Chemistry and Chemical Biology, Harvard University, Cambridge, MA 02138, USA}
\affiliation{Harvard Quantum Initiative, Harvard University, Cambirdge, MA 02138, USA}
\affiliation{Dept. of Physics, University of California, Berkeley, CA 94720, USA}
\email{sajantanand@fas.harvard.edu}
\author{Jack Kemp}
\affiliation{Dept. of Physics, Harvard University, Cambridge, MA 02138, USA}
\affiliation{Cavendish Laboratory, University of Cambridge, Cambridge, CB3 0HE, UK}
\author{Julia Wei}
\affiliation{Dept. of Physics, Harvard University, Cambridge, MA 02138, USA}
\author{Christopher David White}
\affiliation{Center for Computational Materials Science, U.S. Naval Research Laboratory, Washington, D.C. 20375, USA}
\author{Michael P. Zaletel}
\affiliation{Dept. of Physics, University of California, Berkeley, CA 94720, USA}
\affiliation{Material Science Division, Lawrence Berkeley National Laboratory, Berkeley, California 94720, USA}
\author{Norman Y. Yao}
\affiliation{Dept. of Physics, Harvard University, Cambridge, MA 02138, USA}
\affiliation{Harvard Quantum Initiative, Harvard University, Cambirdge, MA 02138, USA}

\begin{abstract}
    Isotropic integrable spin chains such as the Heisenberg model feature superdiffusive spin transport belonging to an as-yet-unidentified dynamical universality class closely related to that of Kardar, Parisi, and Zhang (KPZ).
    To determine whether these results extend to more generic one-dimensional models, particularly those realizable in quantum simulators, we investigate spin and energy transport in non-integrable, long-range Heisenberg models using state-of-the-art tensor network methods.
    Despite the lack of integrability and the asymptotic expectation of diffusion, for power-law models (with exponent $2 < \alpha < \infty$) we observe long-lived $z=3/2$ superdiffusive spin transport and two-point correlators consistent with KPZ scaling functions, up to times $t \sim 10^3/J$.
    We conjecture that this KPZ-like transport is due to the proximity of such power-law-interacting models to the integrable family of Inozemtsev models, which we show to also exhibit KPZ-like spin transport across all interaction ranges.
    Finally, we consider anisotropic spin models naturally realized in Rydberg atom arrays and ultracold polar molecules, demonstrating that a wide range of long-lived, non-diffusive transport can be observed in experimental settings.
\end{abstract}

\maketitle

\begin{figure*}[htbp]
    \centering
\includegraphics{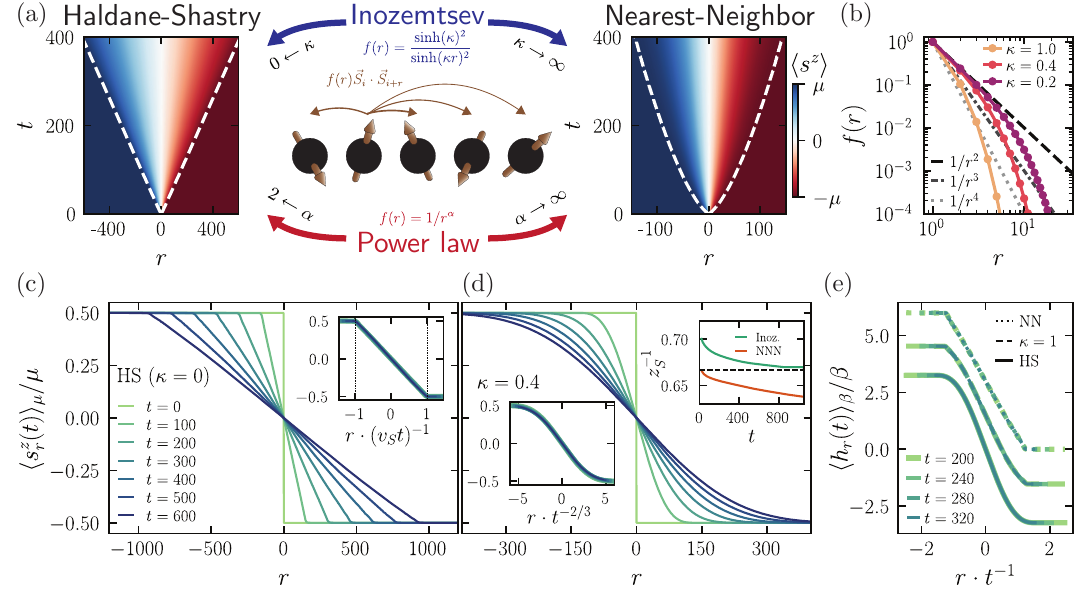}
    \caption{
    \textbf{Landscape of $T=\infty$ transport in integrable, isotropic Heisenberg models.}
    (a) The Haldane-Shastry (HS) and nearest-neighbor (NN) ``fixed points" are integrable, and one can continuously tune between them either by the integrable Inozmentsev models parameterized by constant $\kappa \geq 0$ or generic power-laws with decay $\alpha \geq 2$. The heat plots show the ballistic and superdiffuive ``melting" of the spin domain wall (Eq.~\ref{main-eq:DW}) for the HS and NN models, respectively. The dotted, white line indicates when the spin has deviated $1\%$ from its initial value, with $r \sim t$ and $r \sim t^{2/3}$ for HS and NN.
    (b) Interaction strengths for HS, Inozemtsev, and power-law-interacting models. 
    (c) The linear magnetization profile in the HS model shows ballistic transport and collapses (inset) when rescaled by time $t$, indicating a dynamical critical exponent of $z_S = 1$. Data is rescaled by the spinon speed $v_S = \pi/2$, the speed at which the front propagates~\cite{Bulchandani_2024}. 
    (d) Magnetization profile in Inozemtsev model with $\kappa=0.4$ and collapsed data (left inset) display superdiffusive transport with $z_S = 3/2$. For the Inozemtsev model, $z_S$ calculated from the polarization transfer is stable while that of the comparable non-integrable, next-nearest-neighbor model with coupling $J_2 \approx 0.21$ trends towards diffusion with $z_S^{-1}=1/2$ (right inset).
    (e) Energy profile for HS, $\kappa=1.0$ Inozemtsev, and NN models, rescaled by time $t$ show ballistic transport and dependence on $\kappa$. Profiles have been vertically shifted for visual clarity.
    }
    \label{main-fig:fig1}
\end{figure*}

The emergence of irreversible classical hydrodynamics from charge-conserving, unitary quantum dynamics represents one of the central tenets of statistical physics. 
At high temperatures, conserved charges in a generic system are expected to equilibrate diffusively, resembling the motion of a random walk.
Exceptions to diffusion typically rely on additional structure, such as subdiffusion in localized~\cite{Gopalakrishnan_2020, PhysRevB.105.L140201} models, superdiffusion emerging from Lévy flights~\cite{Schuckert_2020,Joshi_2022} or nodal interactions~\cite{wang2025superdiffusivetransportchaoticquantum}, and anomalous -- most robustly subdiffusive -- transport arising from kinetic constraints~\cite{Morningstar_2020,Feldmeier_2020,Singh_2021,McCarthy_2025,Ljubotina_2023,Chen_2024}.
Recent excitement has focused on a notable exception to diffusion:  infinite-temperature superdiffusive spin transport in the integrable, nearest-neighbor (NN) quantum Heisenberg chain~\cite{Ljubotina_2017,Ljubotina_2019}.
There, some but not all aspects of spin transport~\cite{Ljubotina_2019,Takeuchi_2025} are precisely governed by the Kardar-Parisi-Zhang universality class~\cite{PhysRevLett.56.889,corwin2011kardarparisizhangequationuniversalityclass,Takeuchi_2018}.
%
Combined with extensive numerical explorations~\cite{Dupont_2020,Ilievski_2021,Ye_2022} and seminal experiments~\cite{Scheie_2021,Wei_2022,Rosenberg_2024}, this has led to the conjecture that local, integrable models with a continuous non-Abelian symmetry will generically exhibit KPZ-like superdiffusion with dynamical critical exponent $z=3/2$~\cite{Ilievski_2018,De_Nardis_2020,Ilievski_2021}.
A counterexample to this conjecture is provided by the celebrated Haldane-Shastry (HS) model~\cite{Haldane_1988,Shastry_1988} -- the prototypical long-range-interacting, integrable spin chain -- where both energy and spin transport are ballistic (Fig.~\ref{main-fig:fig1})~\cite{Sirker_2011,Bulchandani_2024}.

This naturally raises the question:  Do longer range interactions immediately preclude the observation of KPZ-like dynamics?
From an experimental perspective, this question is particularly relevant as isolated quantum simulators capable of probing such emergent hydrodynamics (e.g.~Rydberg atom arrays, trapped atomic ions, polar molecules, etc.) often exhibit long-range interactions. 
This allows us to recast our original question:  Does KPZ-like transport ever control the physics of experiments away from fine-tuned, integrable fixed points?

In this Letter, we answer both questions by interpolating between the two aforementioned integrable ``fixed points", the NN Heisenberg spin chain and the $1/r^2$-interacting HS model (Fig.~\ref{main-fig:fig1}).
We interpolate via two routes: (i) non-integrable, power-law-interacting $1/r^\alpha$ Heisenberg models~\cite{Mierzejewski_2023,PhysRevB.110.014308} and (ii) the integrable Inozemtsev family of models~\cite{Inozemtsev_1990}.
For both routes, we investigate infinite temperature spin and energy transport, using tensor network algorithms to explore large systems (up to $L \sim 2000$) and late times (up to $t \sim 1000/J$).

Our main results are threefold.
First, we demonstrate that power-law-interacting Heisenberg models, despite being non-integrable, defy diffusive expectations and exhibit KPZ-like spin dynamics up to the latest time-scales numerically simulable.
Although one expects such anomalous spin transport to ultimately cross over to diffusion, our results suggest that a multitude of experimental platforms realizing such long-range Heisenberg interactions are governed by KPZ-like hydrodynamics. 
Second, we provide a microscopic explanation for our observations above by demonstrating that many extended range Heisenberg models, including those with power-law interactions, can be viewed as a isotropic perturbation to an integrable Inozemtsev model; recent classical and quantum numerics has argued that similarly perturbated locally-interacting integrable models display exceptionally long crossovers to diffusion~\cite{McCarthy_2024,McRoberts_2024,Wang_2025}.
Supported by a numerical investigation of scaling functions, we conjecture that spin transport in the Inozemtsev models, regardless of interaction range, falls into the same KPZ-like universality class as the NN Heisenberg model.
%
%
Finally, motivated by experiment, we explore spin and energy transport in long-range interacting, anisotropic XXZ models.
Although one again expects a crossover to diffusion at infinite times, we observe spin dynamics ranging from subdiffusive to ballistic transport, tuned by the Ising anisotropy.
These behaviors persist to the latest numerically accessible timescales, suggesting that such physics should be readily observable in experiment.

\emph{Models and setup} -- Consider a 1D chain of $N$ spin-1/2 particles with open boundary conditions.
A generic \texttt{SU(2)}-symmetric Heisenberg-like model is given by the following, where where $\vec{S}_i=\left\{ \hat{s}^x_i, \hat{s}^y_i, \hat{s}^z_i \right\}$ are the spin-1/2 operators and $\hat{h}_i$ is the energy density centered on site $i$:
\begin{align}
    \hat{H} = \sum_i \hat{h}_i = \sum_{i} \left(\frac{1}{2}\sum_{j\neq i} f(|i-j|) \hat{\vec{S}}_i \cdot \hat{\vec{S}}_{j} \right).
\end{align}
For the NN Heisenberg model, $f_\text{NN}(r) = \delta_{r,1}$, while for the HS model~\cite{Haldane_1988,Shastry_1988} $f_\text{HS}(r) =1/r^2$~\cite{HSR}.
As depicted in Fig.~\ref{main-fig:fig1}(a), we investigate two continuous paths between these ``fixed points".
The first is the integrable Inozemtsev family of models~\cite{Inozemtsev_1990} with interactions $f_\text{Inoz}^\kappa(r) = \sinh(\kappa)^2/\sinh(\kappa r)^2$ for $\kappa \in \mathbb{R} \geq 0$~\cite{Klabbers_2016,Klabbers_2022,chalykh2024integrabilityinozemtsevspinchain}.
The limit $\kappa \rightarrow 0$ yields the HS model, while $\kappa \rightarrow \infty$ yields the NN model.
The second path is power-law-interacting models, $f_\text{pl}^\alpha(r) = 1/r^\alpha$, parameterized by a decay exponent $2 < \alpha < \infty$. 
Such models are non-integrable and yield the HS and NN models at $\alpha=2$ and $\alpha=\infty$, respectively.
All models considered, integrable or otherwise, conserve both spin and energy.

To study spin and energy transport at infinite temperature, we evolve a weakly polarized domain-wall initial state~\cite{Ljubotina_2017,Ljubotina_2019,Ye_2022}:
\begin{align}
    \rho_\lambda^q(t = 0) \propto e^{\lambda \hat{R}_q} = e^{\lambda \left( \sum_{i<0} \hat{q}_i - \sum_{i\geq0} \hat{q}_i \right)},
    \label{main-eq:DW}
\end{align}
where $\lambda$ is the ``chemical potential" for conserved charge $\hat{Q} = \sum_i \hat{q}_i$~\cite{SOM}.
%
%
The charge transferred from left half to right under time-evolution $\mathcal{P}_Q(t) = \langle\hat{R}(0) - \hat{R}(t) \rangle_\lambda / \lambda$ scales as $\mathcal{P}_Q(t) \propto t^{1/z_Q}$, where $z_Q$ represents the dynamical critical exponent and $\langle \cdot \rangle_\lambda = \mathrm{Tr}(\cdot \rho_\lambda^q(t)) / \mathrm{Tr}(\rho_\lambda^q(t))$.

The exponent $z_Q$ alone is not sufficient to identify the dynamical universality class.
A more sensitive, yet still incomplete, probe is the late-time two-point structure factor, which when rescaled by $t^{1/z_Q}$ collapses to a scaling function characteristic of the universality class:
\begin{align}
    \langle \hat{q}_r(t) \hat{q}_0(0) \rangle = bt^{-1/z_Q} f\left( \frac{br}{t^{1/z_Q}} \right),
    \label{main-eq:SF}
\end{align}
\noindent where $b$ is a model-dependent constant.
Given a domain wall setup, the structure factor can be determined from the spatial gradient of the charge profile, $\Delta \langle q_r \rangle_\lambda = \langle q_{r-1} \rangle_\lambda - \langle q_{r} \rangle_\lambda = 2 \lambda \langle \hat{q}_r(t) \hat{q}_0(0) \rangle_{T=\infty} + \mathcal{O}(\lambda^2)$~\cite{Ljubotina_2019}.
Thus, we choose $\lambda \ll 1$ so that to leading order we study infinite temperature charge dynamics~\cite{SOM}.
Similarly, the expectation value of the current density operator, $j_r^Q(t)$, is also described by a scaling function related to $f$ and is another signature of the dynamical universality class~\cite{SOM}.

\emph{Spin transport at the integrable fixed points} -- It is now well-established by numerics~\cite{Ljubotina_2017,Ljubotina_2019}, analytics~\cite{Ilievski_2018,Gopalakrishnan_2019,De_Nardis_2019,Gopalakrishnan_2019b,De_Nardis_2020,Bulchandani_2020}, and experiments~\cite{Scheie_2021,Wei_2022,Rosenberg_2024} that spin transport in the NN model is superdiffusive with spin dynamical critical exponent $z_{S}=3/2$ (right, Fig.~\ref{main-fig:fig1}a).
Additionally, the spin structure factor, current density, and several other two-point functions are in excellent agreement with that of the KPZ universality class, upon identifying $\hat{s}^z_i \leftrightarrow \partial_x h(x_i)$, where $h(x)$ is the height field in the KPZ equation~\cite{Ljubotina_2019,Takeuchi_2025,MOM}.

While the HS model is known to exhibit ballistic spin transport, its long-range interaction poses a challenge for direct numerical explorations of its hydrodynamic spin transport, necessitating large system sizes and timescales
%
%
To date, spin transport in the HS model has only been numerically explored via exact diagonalization, limiting $N\sim 20$~\cite{Mierzejewski_2023}.
Thus, we develop a state-of-the-art numerical method to time evolve the domain-wall initial state and compute the system's transport properties.
In particular, we treat the mixed state, $\rho_\lambda^q$, as a pure state in a doubled Hilbert space and use matrix product state (MPS) algorithms which are naturally suited to the 1D geometries we study~\cite{Verstraete_2004,Zwolak_2004}.
The long-range interaction specified by $f(r)$ is approximated as $f'(r)$ given by the sum of at least four exponential terms and thus represented as a finite bond dimension matrix product operator.
Then, we utilize the MPS formulation of the time-dependent variational principle, with simulation bond dimension up to $\chi=384$ and largest reliable $t$ used in our analysis determined from the convergence of the charge transfer $\mathcal{P}_q(t)$ within 1\%, to perform time evolution with long-range interactions~\cite{Haegeman_2011,Haegeman_2016,Paeckel_2019,Kloss_2019,SOM}.
Additionally, we impose \emph{weak} $S^z$ symmetry on the evolved mixed state, which provides a significant speedup~\cite{SOM}.
This enables us to study spin transport in the HS model for system sizes up to $N=2000$ and timescales up to $t \sim 600/J$.
Further details and extensive benchmarking of our numerical method can be found in the supplemental~\cite{SOM}.

Upon evolving a weak domain wall, we observe a linearly propagating magnetization profile [Fig.~\ref{main-fig:fig1}(a,c)] that collapses when rescaled by $t$, confirming ballistic spin transport with $z_S = 1$ [Fig.~\ref{main-fig:fig1}(c)].
As a more stringent test of our numerics, we identify the speed of spinon quasiparticles, $v_S = 1.567$, as the velocity with which the front propagates [Fig.~\ref{main-fig:fig1}(a), left], in excellent agreement with recent generalized hydrodynamics (GHD) calculations, which predict, $v_S = \pi/2 \approx 1.571$~\cite{Bulchandani_2024}.
We believe that the discrepancy in the third digit is due to the finite-exponential approximation of the $1/r^2$ interaction~\cite{SOM}.
Our simulations of the spin structure factor and current density are also in excellent agreement with GHD, which provides additional credence to our numerical simulations~\cite{SOM}.

\begin{figure}
    \centering
    \vspace{-2mm}
    \includegraphics{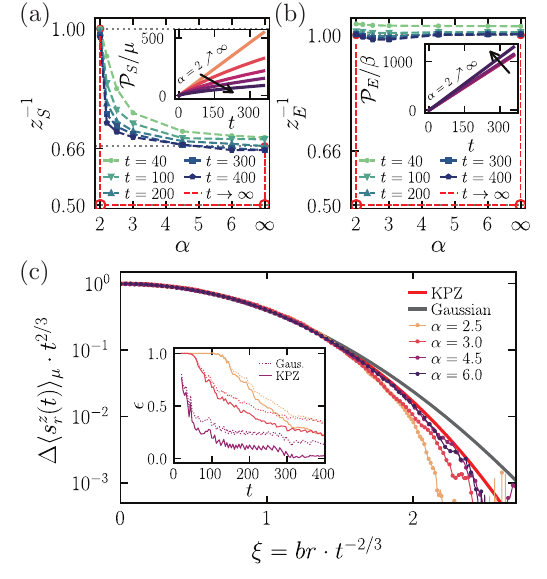}
    \caption{
    \textbf{Energy and spin transport in non-integrable power-law models.}
    Dynamical critical exponent for spin (a) and energy (b) transport as a function of exponent $\alpha$. Dotted red lines are expectations at $t=\infty$.
    (inset) Polarization (a) and energy (b) transfer for $\alpha=2, 2.2, 2.5, 3, \infty$; darker lines correspond to increasing $\alpha$.
    (c) Spin structure factor as a function of $\alpha$ at $t=400$. (inset) Relative error $\epsilon$ between numerically calculated spin structure factor and Gaussian or KPZ predictions for $\alpha=2.5,3,4.5$ at fixed $\xi=1.75$. 
    }
    \label{main-fig:fig2}
\end{figure}

\emph{Hydrodynamics in power-law interacting spin chains} -- Having confirmed that the KPZ-like dynamics of the NN model are not stable to the addition of $1/r^2$ interactions of the HS model, we now ask the broader question: Is this lack of KPZ-like transport a more general feature of long-range interacting systems?
The answer to this question has immediate implications for the dynamics of a wide range of quantum simulation platforms, ranging from trapped atomic ions ($0 < \alpha \leq 3$) to Rydberg atom arrays ($\alpha =3, 6$) and ultracold polar molecules ($\alpha =3$)~\cite{Porras_2004,Deng_2005,Gra__2014,Arrazola_2016,Bermudez_2017,Birnkammer_2022,Bornet_2023,Kranzl_2023,Bornet_2024}.

As generic power-law interacting Heisenberg models are non-integrable, one expects that transport will be diffusive and described by Gaussian two-point functions.
%
%
Surprisingly, for power-law interactions with $\alpha= 2.2, 2.5, 3, 4.5, 6$, we find superdiffusive spin transport and ballistic energy transport up to the latest timescales explored, $t \sim 400 / J$~[Fig.~\ref{main-fig:fig2}(a,b)].
For conserved charge $Q$, we define the instantaneous dynamical scaling exponent $z_Q^{-1}(t) = \mathrm{d} \log(\mathcal{P}_Q(t)) / \mathrm{d} \log(t)$, evaluated numerically from a linear fit over a window $\Delta t = 20$.
For $2 <\alpha \leq 3$, $z_S^{-1}(t)$ exhibits significant finite-time flow toward smaller values, while for $\alpha \gtrsim 4$, this flow is considerably weaker~[Fig.~\ref{main-fig:fig2}(a) and inset].
Nevertheless, across all models, the behavior is consistent with $z_S^{-1}(t)$ approaching a finite-time saturation value of $2/3$, in agreement with that of the integrable NN model.
This integrable-like behavior of power-law models is even more pronounced in the energy transport, which exhibits a ballistic $z_E^{-1} = 1$, independent of $\alpha$~[Fig.~\ref{main-fig:fig2}(b)].

While suggestive, the spin dynamical critical exponent alone is not sufficient to diagnose KPZ-like transport.
To this end, we measure the two-point spin structure factor [Fig.~\ref{main-fig:fig2}(c)] at $t=400/J$ and find excellent agreement with the KPZ scaling function for $\alpha \gtrsim 4$.
For smaller values of $\alpha$, where the spin dynamical critical exponent still flows over these timescales, agreement with the KPZ scaling function improves with increasing time, as one sees from the relative error with respect to the KPZ and Gaussian scaling functions (inset)~\cite{SOM}.
Finally, the fact that we do not observe any evidence of the expected crossover to diffusion for both spin and energy transport indicates that the intermediate time integrable-like dynamics of such power-law-interacting models is surprisingly robust.

\begin{figure}
    \vspace{-2mm}
    \centering
    \includegraphics{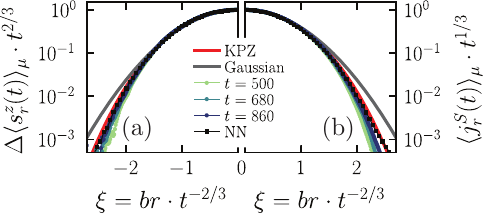}
    \caption{
    \textbf{$T=\infty$ Inozemtsev spin structure factor and current density.}
    (a) Spin structure factor and (b) spin current density for $\kappa=0.4$ Inozemtsev model show excellent agreement with KPZ predictions~\cite{Pr_hofer_2004} and those of the NN model and \emph{not} with Gaussian predictions.
    }
    \label{main-fig:fig3}
\end{figure}

\emph{Proximity to the Inozemtsev family of models} -- We conjecture that the origin of this anomalous transport is the proximity of the power-law models to the integrable Inozemtsev family of models [Fig.~\ref{main-fig:fig1}].
Indeed, \emph{any} SU(2) symmetric Heisenberg-like model with interaction $f(r)$ can be viewed as a symmetry preserving perturbation, $f'(r)$, to an Inozemtsev model with coupling $\kappa$: $f' = f - f_\text{Inoz}^\kappa$.
Crucially, recent numerical studies of SU(2) symmetry-preserving, integrability-breaking perturbations have suggested extremely long crossover timescales $\sim 1 / \epsilon^{\nu}$, where $\epsilon = ||f'||_1$ characterizes the perturbation strength and $3 \leq \nu \leq 8$~\cite{McCarthy_2024,McRoberts_2024,Wang_2025}.
By treating Inozemtsev $\kappa$ as a variational parameter and minimizing $\epsilon = ||f'||_1$ for each power-law $alpha$, we find that power-law-models can be viewed as \emph{weak} perturbations to an Inozemtsev model of strength at most $\epsilon \lesssim 0.098$, peaking for $\alpha=2.3$. 
Thus even with positional disorder, we expect integrable features of Inozemtsev models to control the transport of power laws to anomalously long timescales~\cite{SOM}.
For perturbations to the HS model, however, we find a crossover from ballistic spin transport to superdiffusion with timescale scaling with exponent $\nu=2$, again highlighting the peculiarity of this model among long-range-interacting, integrable spin chains~\cite{SOM}.

We now probe spin transport in Inozemtsev models, focusing on $\kappa=0.4$ as a representative example.
We extract $z_S$ from the collapse of the polarization profile [Fig.~\ref{main-fig:fig1}(d)], finding $z_S = 3/2$.
Next, we compute both the two-point spin structure factor [Fig.~\ref{main-fig:fig3}(a)] and the current density [Fig.~\ref{main-fig:fig3}(b)], observing excellent agreement with the KPZ scaling functions.
More generally, KPZ-like spin transport is found for all models investigated, $0.1 \leq \kappa \leq 2.0$ (see Supplemental Materials~\cite{SOM}); we note that for smaller $\kappa$, there is an extended crossover from ballistic behavior, likely due to proximity to the HS model~\cite{PLP}.

A few remarks are in order.
First, in contrast to the family of power-law interacting models, since the Inozemtsev models are themselves integrable, we expect the KPZ-like spin transport behavior to persist to infinite times~\cite{HOM}.
Second, to determine whether the KPZ-like behavior of the power-law models simply owes to proximity to the NN model, we truncate the $\kappa=0.4$ Inozemtsev model to next-nearest-neighbor, $J_2 \approx 0.21$ (with perturbation strength $\epsilon \approx 0.126$), and study spin transport.
One observes a slow crossover towards diffusion, demonstrating that the long-range Inozemtsev tails are important for stabilizing KPZ-like dynamics [right inset, Fig.~\ref{main-fig:fig1}(d)].
Finally, we also explore energy transport in the Inozemtsev modelsand find ballistic transport as indicated by the collapse of the energy profile when rescaled by time [Fig.~\ref{main-fig:fig1}(e)].
However, we find that the \emph{shape} of the profile depends on $\kappa$ and thus the range of interaction.
Similarly, the energy structure factor and current density also exhibit $\kappa$-dependence and do not collapse onto a single curve~\cite{SOM}.
This suggests that energy transport in the Inozemtsev family cannot be identified with a single dynamical universality class and instead smoothly interpolates between that of the HS and NN models.

\begin{figure}
    \vspace{-2mm}
    \centering
    \includegraphics{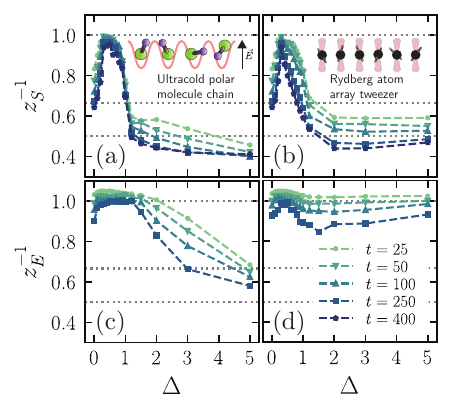}
    \caption{
    \textbf{Energy and spin transport in cold atom simulator models.}
    Spin (a,b) and energy (c,d) transport for the polar molecule $H_\text{pm}$ (a,c) (Eq.~\ref{main-eq:pm}) and dipolar Rydberg $H_\text{dR}$ (b,d) (Eq.~\ref{main-eq:dR}) models.
    }
    \label{main-fig:fig4}
\end{figure}

\emph{Spin and energy transport in generic quantum simulators} -- Thus far, we have demonstrated that KPZ-like spin transport can be observed in long-range interacting Heisenberg models, even when integrability is broken.
%
%
Now, we broaden our scope and ask whether non-diffusive hydrodynamics can also be observed in long-range-interacting, \emph{anisotropic} spin models, which are neither integrable nor SU(2) symmetric. 

Motivated by experiment, we focus on two such models: (i) the dipolar XXZ model native to polar molecules and magnetic atoms and (ii) a mixed-power-law XXZ model native to Rydberg atom arrays~\cite{Nguyen_2018,Scholl_2022,Bornet_2023, chen2024spectroscopyelementaryexcitationsquench,Bornet_2024,Emperauger_2025,kunimi2025proposalrealizingheisenbergtypequantumspin}. 
Let us begin with the dipolar XXZ model, whose Hamiltonian is given by: 
\begin{align}
    H_\text{dip} = \sum_{i} \frac{1}{r^3} \left[ \hat{s}^x_i \cdot \hat{s}^x_{i+r} + \hat{s}^y_i \cdot \hat{s}^y_{i+r} + \Delta \hat{s}^z_i \cdot \hat{s}^z_{i+r} \right],
    \label{main-eq:pm}
\end{align}
where $\Delta$ characterizes the anisotropy.
Although diffusion is again expected for both spin and energy transport as $t \rightarrow \infty$, for all $\Delta$ we find that any such diffusion must set in at an anomalously long time scale.
In particular, up to the latest timescales explored ($t \sim 400/J$), we observe two ``regimes'' of spin transport [Fig.~\ref{main-fig:fig4}(a)]:  (i) ballistic  transport for $0.3 \lesssim \Delta  \lesssim 0.5$ and (ii) subdiffusion for $\Delta \geq 2$.
The former is consistent with recent numerical and analytic predictions of a long-lived, approximately conserved spin-current~\cite{Mierzejewski_2023,song2025theoryquasiballisticspintransport}, while the latter is reminiscent of recent results in the perturbed NN XXZ model~\cite{De_Nardis_2022}.
For the energy transport [Fig.~\ref{main-fig:fig4}(c)], we again observe ballistic behavior at small $\Delta$, albeit over a significantly larger range of anisotropies, $0.1 \lesssim \Delta \lesssim 1.2$; at larger $\Delta$, the system seems to be slowly converging toward the diffusive expectation. 

Finally, we turn to the mixed-power-law model with Hamiltonian:
\begin{align}
    H_\text{ryd} = \sum_{i} \frac{1}{r^3} \left[ \hat{s}^x_i \cdot \hat{s}^x_{i+r} + \hat{s}^y_i \cdot \hat{s}^y_{i+r}\right] + \frac{\Delta}{r^6} \hat{s}^z_i \cdot \hat{s}^z_{i+r}.
    \label{main-eq:dR}
\end{align}
Apart from a narrow region of ballistic transport near $\Delta=0.3$, spin transport appears to be more conventionally diffusive.
Energy transport, however, is highly anomalous and is superdiffusive for all investigated $\Delta$.
This is in sharp contrast to the flow toward diffusion for the large $\Delta$ region of $H_\text{dip}$.

\emph{Outlook} -- %
%
Our work opens several directions for future study. 
First, generalized hydrodynamics for Inozemtsev models~\cite{Bertini_2016,Castro_Alvaredo_2016} may enable analytical characterization of the similarities to and differences from the HS and NN limits. 
Second, deformed HS and Inozemtsev models~\cite{Lamers_2018,klabbers2025landscapesintegrablelongrangespin,Klabbers_2024,moussa2024qdeformedhaldaneshastrychainqi}, which break SU(2) to U(1) while preserving integrability, allow one to ask whether ballistic transport and diffusion emerges in the easy-plane and easy-axis regime, respectively, analogous to the NN XXZ model. 
Finally, it would be interesting to test whether finite temperature can restore signatures of anomalous transport in non-integrable quantum models, as recently reported in simulations of classical field theories~\cite{koterle2026anomaloustransportnonintegrableclassical}.

\begin{acknowledgements}
\textbf{Acknowledgments:}
We thank Vir Bulchandani, Michele Fava, Sarang Gopalakrishnan, Joel Moore, Siddharth A. Parameswaran, Romain Vasseur, and Zack Weinstein for helpful discussions.
SA thanks Johannes Hauschild for discussions on tensor network methodology and the TeNPy library which was used for tensor network calculations~\cite{Hauschild_2018,Hauschild_2024}.
S.A. and MPZ were supported by the U.S. Department of Energy, Office of Science, Basic Energy Sciences, under Early Career Award No. DE-SC0022716.
This work was supported by the U.S. Department of Energy via the Office of Science, National Quantum Information Science Research Centers, Quantum Systems Accelerator and via the QuantISED program (Geoflow collaboration) as well as the Army Research Office via grant no. W911NF-24-1-0079.
Computing resources were provided by (1) the National Energy Research Scientific Computing Center (NERSC), a U.S. Department of Energy Office of Science User Facility located at Lawrence Berkeley National Laboratory, operated under Contract No. DE-AC02-05CH11231 using NERSC Award No. BES-ERCAP0024710, (2) the Lawrencium computational cluster resource provided by the IT Division at the Lawrence Berkeley National Laboratory (Supported by the Director, Office of Science, Office of Basic Energy Sciences, of the U.S. Department of Energy under Contract No. DE-AC02-05CH11231), and (3) the Texas A\&M University High Performance Research Computing through allocation PHY250200 from the Advanced Cyberinfrastructure Coordination Ecosystem: Services and Support (ACCESS) program, which is supported by National Science Foundation grants 2138259, 2138286,2138307, 2137603, and 2138296.
S.A.~also acknowledges support from the Harvard Quantum Initiative and the Center for Ultracold atoms (an NSF Physics Frontier center).
J.K. acknowledges support from EPSRC Grant No. EP/V062654/1. 
J.W. acknowledges support from the Department of Energy Computational Science Graduate Fellowship under award number DE-SC0022158. 
CDW gratefully acknowledges ARO grant W911NF-23-1-0242, ARO grant W911NF-23-1-0258, and NSF QLCI grant OMA-2120757; the work was completed while C.D.W. held an NRC Research Associateship award at the United States Naval Research Laboratory. 
N.Y.Y. acknowledges support from a Simons Investigator award. 
\end{acknowledgements}

\bibliographystyle{apsrev4-1} 
\bibliography{refs} 

\begin{thebibliography}{84}
\bibitem [{HSR()}]{HSR}%
  \BibitemOpen
  \href@noop {} {}\bibinfo {note} {The model was originally defined for spin-1/2 particles on a ring, where the interaction strength is a function of the chord distance on the circle. Here, we take the $L \rightarrow \infty$ to get an infinite chain and the $1 / r^2$ interaction. While the open-chain model at any finite $L$ is technically not integrable due to boundary terms in the Hamiltonian, these do not affect our observations due to the large system sizes considered.}\BibitemShut {Stop}%
\bibitem [{PLP()}]{PLP}%
  \BibitemOpen
  \href@noop {} {}\bibinfo {note} {Just as power-law models can be viewed as SU(2) preserving but integrability-breaking perturbations to an Inozemtsev model, Inozemtsev models can be viewed as SU(2) and integrability preserving perturbations to the HS model.}\BibitemShut {Stop}%
\end{thebibliography}

\end{document}


\title{Supplemental material for ``Robustness of Kardar-Parisi-Zhang-like transport in long-range interacting quantum spin chains''}

\author{Sajant Anand}
\affiliation{Dept. of Physics, Harvard University, Cambridge, MA 02138, USA}
\affiliation{Dept. of Chemistry and Cheical Biology, Harvard University, Cambridge, MA 02138, USA}
\affiliation{Harvard Quantum Initiative, Harvard University, Cambirdge, MA 02138, USA}
\affiliation{Dept. of Physics, University of California, Berkeley, CA 94720, USA}
\email{sajantanand@fas.harvard.edu}
\author{Jack Kemp}
\affiliation{Dept. of Physics, Harvard University, Cambridge, MA 02138, USA}
\affiliation{Cavendish Laboratory, University of Cambridge, Cambridge, CB3 0HE, UK}
\author{Julia Wei}
\affiliation{Dept. of Physics, Harvard University, Cambridge, MA 02138, USA}
\author{Christopher David White}
\affiliation{Center for Computational Materials Science, U.S. Naval Research Laboratory, Washington, D.C. 20375, USA}
\author{Michael P. Zaletel}
\affiliation{Dept. of Physics, University of California, Berkeley, CA 94720, USA}
\affiliation{Material Science Division, Lawrence Berkeley National Laboratory, Berkeley, California 94720, USA}
\author{Norman Y. Yao}
\affiliation{Dept. of Physics, Harvard University, Cambridge, MA 02138, USA}
\affiliation{Harvard Quantum Initiative, Harvard University, Cambirdge, MA 02138, USA}

\date{\today}
\maketitle

\tableofcontents

\section{Numerical setup}
\label{supp-sec:setup}
First we describe our numerical setup in detail, covering the construction of the initial domain wall state, time evolution with the time-dependent variational principle (TDVP) in the doubled Hilbert space, charge conservation in the doubled Hilbert space, and representing long range Hamiltonians as the sum of exponentials. We also provide detailed benchmarking, showing that our results are converged in various hyperparameters of the algorithm, such as bond dimension, TDVP step size, truncation precision, and specifics of initial state construction.
Together, this enables the accurate study of quantum transport in up to $N \sim 2000$ spins and times $t \sim 1000 / J$.

\subsection{Structure of the initial state}
\label{supp-sec:initial_state}
Time evolution of a weak domain wall (DW), given by
\begin{align}
    \rho_\lambda^q(t = 0) \propto e^{\lambda \hat{R}_q} = e^{\lambda \left( \sum_{i<0} \hat{q}_i - \sum_{i\geq0} \hat{q}_i \right)},
    \label{supp-eq:DW}
\end{align}
is a numerically effective method for probing infinite temperature hydrodynamics in systems with a conserved quantity~\cite{Ljubotina_2017,Ljubotina_2019,Ye_2022}.
The state is defined in terms of a charge density $\hat{q}_i$ centered on site $i$ and a potential $\lambda$.
The weak DW is a physically motivated state with an excess and deficit of charge on the left and right of the system; intuitively, we are performing a weak measurement on the infinite temperature density matrix to bias the charge oppositely in the two halves of the chain.
For spin transport with $\hat{q}_i = \hat{s}_i^z$, the initial state $\rho_\lambda^{S}(t = 0)$ has an excess and a deficit of magnetization on the left and right halves, respectively.
For models with an extensive energy (the energy grows linearly with systems size but not faster; i.e. the interactions are not so long ranged that the interactions are diverging), one can probe energy transport with $\hat{q}_i = \hat{h}_i$, where $\hat{h}_i$ is the energy density centered on site $i$; in that case the potential is inverse temperature $\beta$, and initial state $\rho_\beta^{E}(t = 0)$ represents two systems at inverse temperatures $\beta$ and $-\beta$ joined end-to-end.

Time-evolving the DW state naturally allows one to measure the charge transfer between the two halves as a function of time, given by
\begin{align}
    \mathcal{P}_Q(t) =  \sum_{i \geq 0} \langle \hat{q}_i(t) - \hat{q}_i(0) \rangle \sim t^{1/z_Q}
    \label{supp-eq:PT}
\end{align}
The charge transfer allows us to determine the dynamical critical exponent $z_Q$ for charge $\hat{Q}$.
In addition, this state is natural to prepare in experiment and has been used in a quantum gas microscope study of spin transport of the nearest-neighbor XXZ model in both one and two dimensions~\cite{Wei_2022}.
Moreover, measurements of the time evolved domain wall provide access to the infinite temperature charge structure factor, up to corrections at second order in the charge potential, from the spatial gradient of the local charge density~\cite{Ljubotina_2019}.
\begin{align}
    \Delta \langle \hat{q}_i(t) \rangle_{\lambda} &= \langle \hat{q}_{i-1}(t) \rangle_\lambda - \langle \hat{q}_i(t) \rangle_\lambda  \nonumber \\
    &\approx 2 \lambda \langle \hat{q}_i(t) \hat{q}_0(0) \rangle_{T = \infty} - \lambda \langle \hat{q}_i(t) \hat{q}_{-\infty}(0) \rangle_{T = \infty} - \lambda \langle \hat{q}_i(t) \hat{q}_{\infty}(0) \rangle_{T = \infty} + \mathcal{O}(\lambda^2),
    \label{supp-eq:gradient}
\end{align}
where $\langle \cdot \rangle_\lambda$ indicates expectation values in the weak domain wall state with domain wall strength $\lambda$ and $\langle \cdot \rangle_{T = \infty} = \Tr(\cdot) / 2^N$ is the infinite temperature expectation value.
The final two terms in Eq.~\eqref{supp-eq:gradient} reduce to the product $\langle \hat{q}_0 \rangle_{T = \infty}^2 $ as the infinite temperature state is stationary under time evolution, translationally invariant, and finitely correlated (here correlation length is 0).
Thus the spatial gradient gives the connected correlation function, although typically the later two terms will be identically zero at infinite temperature as operator $q$ is often traceless.
We will see in Sec.~\ref{supp-sec:Drude} with a different choice of domain wall initial state that the 1-point functions can be nonzero.

\textbf{Spin domain wall:} For a charge where the local density is onsite, such as spin with charge $\hat{s}_i^z$, the domain wall density matrix factorizes into an independent density matrix on each site, $\rho = \otimes_i^{L} \rho_i$.
Supposing we are on the left half of the chain with $i < 0$, the local state is
\begin{align}
    \rho_i = \frac{e^{\lambda \hat{s}_i^z}}{\mathrm{Tr}(e^{\lambda \hat{s}_i^z})} = \frac{\mathbb{I}_i^{2 \times 2}}{2} + \tanh\left(\frac{\lambda}{2}\right) \hat{s}_i^z,
    \label{supp-eq:spin_DW}
\end{align}
where $\mathbb{I}^{2 \times 2}$ is the $2 \times 2$ identity matrix.
We have normalized this density matrix so that it has trace 1.
We define $\mu = \tanh\left(\frac{\lambda}{2}\right)$ so that
\begin{align}
    \rho = \bigg(\mathbb{I}_i^{2 \times 2}/2 + \mu \hat{s}_i^z\bigg)^{\otimes N//2} \bigg(\mathbb{I}_i^{2 \times 2}/2 - \mu \hat{s}_i^z\bigg)^{\otimes (N-N//2)}
    \label{supp-eq:actual_spin_DW}
\end{align}
and $\langle \hat{s}_i^z\rangle_\rho = \pm \mu/2$.
This domain wall state breaks SU(2) symmetry since spin orientation along $\hat{z}$ is specified, and thus transport from this initial state is expected in the $t \rightarrow \infty$ limit to be diffusive; however, the timescale at which this occurs has been numerically demonstrated to be exceedingly long and has not been observed in experiment or numerics for $\mu \ll 1$~\cite{Misguich_2017,Collura_2020,10.21468/SciPostPhys.7.2.025}.
Typically we choose $\mu=0.05$, but as we show in Fig.~\ref{supp-fig:hyperparameters}(c), our results are insensitive to the choice of $\mu$, provided $\mu \ll 1$.

\textbf{Energy domain wall:} For a charge with a multisite and generically non-commuting local density, such as the energy with local density $\hat{h}_i$ that acts on two or more sites, the domain wall initial state will not be the product of density matrices on each site and thus is represented as a matrix product operator (MPO) with non-trivial ($\chi > 1$) bond dimension.
For the isotropic models we consider, the energy density $\hat{h}_i = \frac{1}{2}\sum_{j\neq i} f(|i-j|) \hat{\vec{S}}_i \cdot \hat{\vec{S}}_{j}$ contains half of the interaction between site $i$ and all other sites; anisotropic energy densities are defined analogously.
To represent the initial DW state, we then use the WII method~\cite{Zaletel_2015} to construct a compact MPO approximation to the desired state; if the matrix product operator representation of $\hat{R}_q = \left( \sum_{i<0} \hat{h}_i - \sum_{i\geq0} \hat{h}_i \right)$.
Note that terms coupling sites $(j,k)$ with $j < i$ and $k \geq i $ will cancel, so effectively $\hat{R}_q$ measures the difference in energy between the two disconnected halves of the spin chain and does not introduce any coupling between the two sites.
Effectively we are joining together two disconnected subsystems of size $L/2$, with temperature $-\beta$ and $\beta$, end-to-end; dynamics will then introduce coupling between the subsystems and cause energy to equilibrate.

If $\hat{R}_q$ has a maximum bond dimension $D$, the second order WII approximation of $e^{\lambda \hat{R}_q}$ will have a maximum bond dimension $D+1$~\cite{Zaletel_2015}.
It is known that this approximation will omit a subset of terms of $\rho = e^{\lambda \hat{R}_q}$ starting at order $\mathcal{O}(\lambda^2)$, but supposing $\lambda \ll 1$, these terms can be safely ignored.
In fact, these higher order terms are undesirable, as they contribute to beyond linear response effects in the structure factor, evaluated from expectation values of the time evolved domain wall.
We choose $\beta = 0.01$, but as is the case for $\mu$ in spin transport, our results are insenstive to this, provided $\beta \ll 1$.

\textbf{Linear response domain wall:} The domain wall state as constructed contains terms probing beyond linear response.
For example, the spin domain wall (Eq.~\eqref{supp-eq:spin_DW}) when expanded in terms of Pauli strings contains terms like $\hat{s}_i^z \hat{s}_j^z$ with amplitude $\mathcal{O}(\lambda^2)$. 
To avoid these higher order terms entirely, rather than perturbatively with small $\lambda$, we can construct an initial state $\tilde{\rho} \propto \mathbb{I}^{2^N \times 2^N} + \mu' \hat{R}_q$, which can be represented exactly as an MPO with the same bond dimension as $\hat{R}_q$; here $\mathbb{I}^{2^N}$ is the $N$-spin identity matrix of dimension $2^N \times 2^N$, and the normalization coefficient is $1/2^N$ such that $\tr \tilde{\rho} = 1$.
Generically $\tilde{\rho}$ will not be a positive density matrix and thus is not a physical state, but for spin transport, one can choose $\mu'$ such that the $\tilde{\rho}$ is positive.
Suppose we choose $\mu' > 0$.
The coefficient of the bit string state $\ket{\ldots \downarrow \downarrow \uparrow \uparrow \ldots}\bra{\ldots \downarrow \downarrow \uparrow \uparrow \ldots}$ in $\tilde{\rho}$ will be $1 - \sfrac{\mu' N}{2}$; this is the smallest magnitude coefficient possible as this basis state is anti-aligned with the potential $\mu'$ on each site.
So by choosing $\mu' < \sfrac{2}{(N)}$, we guarantee that $\tilde{\rho}$ is a positive semi-definite density matrix.
Note that $\tilde{\rho}$ is precisely the set of Pauli strings found by truncating $\rho$ (Eq.~\eqref{supp-eq:actual_spin_DW} to first order.
So when choosing $\mu' = \mu$ such that $\Tr(\hat{s}^z_i \tilde{\rho}) = \Tr(\hat{s}^z_i \rho) = \pm \mu/2$, we find no difference between simulations with $\rho$ and $\tilde{\rho}$ (Fig.~\ref{supp-fig:hyperparameters}(c) inset).

\subsection{Time evolution in the doubled Hilbert space}
\label{supp-sec:doubled_HS}
Having constructed our initial mixed state on $N$ spins, we wish to evolve it in the Schr{\"o}dinger picture, $\rho(t) = U(t) \rho(0) U^\dagger(t)$ with $U(t) = e^{-i \hat{H} t}$.
To use conventional matrix product state formalism, we ``vectorize" the density matrix into a pure state on a ``doubled" Hilbert space of $2N$ spins or equivalently $N$ spins with a local Hilbert space of $2^2 = 4$,
\begin{align}
\rho(t) = \sum_{i,j} \rho_{ij}(t) \ket{i}\bra{j} \mapsto | \rho(t) \rangle \rangle = \sum_{i,j} \rho_{ij}(t) \ket{i,j}.
\end{align}
\noindent Operationally, we simply flip the bra to a ket, and the induced Hilbert-Schmidt overlap is $\langle \langle \sigma | \rho \rangle \rangle = \mathrm{Tr}(\sigma^\dagger \rho)$.
We time evolve this vectorized pure state with standard MPS algorithms~\cite{Verstraete_2004,Zwolak_2004,Hauschild_2018,Paeckel_2019}, albeit with modifications (discussed in the next section) to account for the doubling of the Hilbert space.
The MPS representing the density matrix is truncated as usual by discarding singular values on each bond according to either a bond dimension upper bound or precision cutoff for the singular value decomposition (SVD).

Note that truncation generically will not preserve hermiticity ($\rho = \rho^\dag$), positive semi-definiteness ($\rho \geq 0$, which we refer to as positivity), or unit trace of the density matrix ($\Tr(\rho) = 1$), so we are not guaranteed to have a physical density matrix after time evolution.
(Some work, e.g. \cite{whiteQuantumDynamicsThermalizing2018a}, avoids the hermiticity problem
by working with strictly real density matrices in a hermitian bases,
but this is not compatible with charge conservation.)
However, the deviation in trace can be accounted for when calculating expectation values of $\rho(t)$.
Loss of hermiticity manifests in expectation values becoming complex and can be diagnosed by the ratio of the real to complex values.
Loss of positivity is harder to diagnose, as checking positivity of a density matrix is \texttt{NP}-hard and involves diagonalizing the density matrix at cost exponential in $N$.

\begin{figure}
    \centering
    \includegraphics{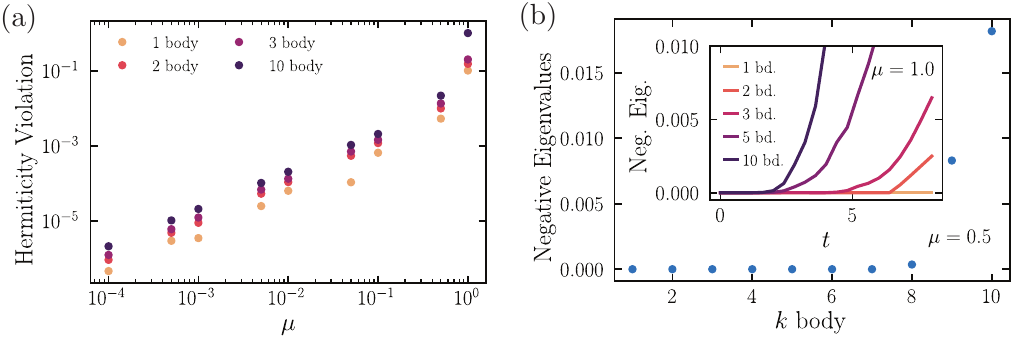}
    \caption{(a) Hermiticity of $k$-body reduced density matrices as a function of domain wall strength for time evolution of $L=10$ nearest-neighbor XXZ model. Data is measured at time $t=80$, and evolution is performed by TDVP with a (uniform) timestep of $dt=0.4$. (b) Total weight of negative eigenvalues for $k$-body reduced density matrices for an evolution of an $L=10$ domain wall with $\mu=0.5$ at time $t-80$. (inset) Total weight of negative eigenvalues as a function of time for various $k$-body reduced density matrices for an evolution of an $L=10$, pure state ($\mu=1.0$) domain wall.
    }
    \label{supp-fig:positivity}
\end{figure}

\textbf{Small test case:} To understand the the loss of hermiticity and positivity, we study the dynamics of an $N=10$ chain of spin-1/2 particles, evolving under the nearest-neighbor Heisenberg model, $H = \sum_i J \vec{S}_i \cdot \vec{S}_{i+1}$, with open boundary conditions.
The system is evolved with TDVP with timestep $\mathrm{d}t=0.4$ and thus is representative of the simulations performed in this work with long-range Hamiltonians.
In Fig.~\ref{supp-fig:positivity}(a), we show the non-hermitian-ness of reduced density matrices (RDMs) of diameter and bodyness $k$ at time $Jt=80$ for spin domain walls of strength $\mu$.
Note that $\mu=1$ corresponds to the pure state DW $\ket{\ldots \downarrow \downarrow \uparrow \uparrow \ldots}\bra{\ldots \downarrow \downarrow \uparrow \uparrow \ldots}$.
We see that non-hermiticity increases with both $k$-bodyness and $\mu$, yet remains small at late times for $\mu \leq 0.1$.

In Fig.~\ref{supp-fig:positivity}(b), we show the total weight of negative eigenvalues, $\sum_{\lambda_i < 0} \lambda_i$ for $\rho = \sum_i \lambda_i \ket{\lambda_i}\bra{\lambda_i}$, for an evolved $\mu=0.5$ state at time $Jt=80$ for increasing diameter RDMs.
Note that we ``hermitian-ize" the RDMs if necessary, $\rho \rightarrow (\rho + \rho^\dagger)/2$.
We find that smaller diameter RDMs remain positive while larger $k=8,9,10$-body operators develop negative weights.
This phenomenon is especially prevalent in the dynamics of a pure state, $\mu=1$ DW (inset of Fig.~\ref{supp-fig:positivity}(b)), where we see that non-positivity manifests earlier in time in the larger RDMs than it does in the smaller RDMs; In fact, the one body RDMs remain positive for the duration of the simulation, while $k=2$ and higher body RDMs develop negative eigenvalues.
However, loss of positivity does not immediately manifest in nonphysical \textit{local} expectation values.
Positivity is a ``non-local" property, as it requires diagonalizing the density matrix and thus rotating away from the computational basis into the eigenbasis.
Thus even when global positivity is lost at time $t^*$, positivity in a local basis, such as the computational basis of $\hat{s}_i^z$, or positivity of small diameter RDMs can still be preserved to later  $t' > t*$.
Positivity in the computational basis means that the distribution over bit strings, given by the diagonal of the density matrix, is a well-defined probability distribution.
Positivity in $k$-body and $k$-diameter RDMs ensures that local expectation values supported on these RDMs remain physical (e.g. within the bounds expected for Pauli operators) for an extended period of time even when $\rho(t>t^*)$ formally is not positive.

Finally, note that for $\mu < 0.5$, all RDMs, including the $10$ body one for the entire system, remain positive for the entirety of the simulation.
A small $\mu$ yields a state closer to the maximally mixed identity matrix and thus maintains positivity longer.
For the simulations in this work with $\mu=0.05$ to approximate $T=\infty$ transport, we consistently find the real parts of expectation values to be at least $10^3$ larger in magnitude than the imaginary part introduced by non-Hermiticity.
Additionally, as we are primarily interested in few-body properties, namely spin and energy charge density and spin and energy current, we see no indications of loss of positivity, such as an expectation value of Pauli operators outside the range $[-1, 1]$.

\subsection{Evolution with non-local Hamiltonian in the doubled Hilbert space}
\label{supp-sec:TDVP}
In this work, we are concerned with evolution by long-range Hamiltonians where in principle two spins interact regardless of how far apart they are separated.
Thus, the time evolving block decimation (TEBD) algorithm, the standard approach for time evolving an MPS under a local Hamiltonian, cannot be efficiently applied~\cite{Vidal_2004,Paeckel_2019}, even with swap gates and grouping of sites.
Instead, we use the MPS formulation of the time-dependent variational principle (TDVP), which operates based on a matrix product operator (MPO) formulation of the Hamiltonian and thus is naturally suited to arbitrary range interactions~\cite{Haegeman_2011,Haegeman_2016}.

To perform TDVP on the vectorized density matrix $|\rho\rangle\rangle$, we thus need the operator that exponentiates to $U \otimes U^* = e^{-i \hat{H} t} \otimes e^{i \hat{H}^* t}$, where when vectorizing $\rho$ we take the transpose of $U^\dag$ acting on the bra legs.
Thus, given a Hamiltonian $\hat{H}$ for the single copy Hilbert space, we wish to construct the doubled Hamiltonian $\hat{H}^D = \hat{H} \otimes \mathbb{I}^{2^N \times 2^N} - \mathbb{I}^{2^N \times 2^N} \otimes \hat{H}^*$ that acts in the doubled Hilbert space.
Note that evolution under this Hamiltonian does not couple the bra and ket copies of the Hilbert space, so the purity of the initial density matrix is preserved up to errors from truncation.
In fact, we could in principle evolve the two sites separately, $\rho \rightarrow U \rho \rightarrow U \rho U^\dagger$.
However, the intermediate state $U \rho$ is not a physical density matrix, and truncation via SVD may fail to keep physically relevant information.

\textbf{Doubled MPO construction:} Given the MPO for $\hat{H}$, we need to construct an MPO for $\hat{H}_D$.
First, we introduce some MPO terminology.
An MPO $O$ for an operator $\hat{O}$ can be written in the standard, upper block triangular form~\cite{Crosswhite_2008,Pirvu_2010,Hubig_2017,Parker_2020}
\begin{align}
    O &= \bra{v_L} W^{[0]} W^{[1]} \ldots W^{[N-1]} \ket{v_R}, \\
    W^{[i]}_{\alpha \beta} &= \left[ \begin{array}{ccc} \mathbb{I}^{2 \times 2} & C^{[i]} & D^{[i]} \\ 0 & A^{[i]} & B^{[i]} \\ 0 & 0 & \mathbb{I}^{2 \times 2} \end{array}\right].
    \label{supp-eq:MPO_standard_form}
\end{align}
We suppress the physical indices $\sigma, \sigma'$ of the tensor four-dimensional tensor $W^{[i]}$ and treat it as a matrix in the virtual space where the entries are operators acting on the physical Hilbert space.
The boundary vectors $\ket{v_L}$ and $\ket{v_R}$ are standard for finite operators and MPOs.
\begin{align}
    \ket{v_L} &= \left[ \begin{array}{ccc} \mathbb{I}^{2 \times 2} & 0 & 0 \end{array} \right]^\top \\
    \ket{v_R} &= \left[ \begin{array}{ccc} 0 & 0 & \mathbb{I}^{2 \times 2} \end{array} \right]^\top
\end{align}
The operator can be viewed as a sum of operator strings in a chosen operator basis, typically taken to be the Pauli or spin-$\sfrac{1}{2}$ operator basis.
In this picture, the blocks of $W$ can be understood in terms of the operator strings they generate as follows, where we suppress the site index $i$:
\begin{itemize}
    \item $D$ - onsite terms acting only on site $i$.
    \item $C$ - terms where the leftmost operator is on site $i$, i.e. multisite terms that start on site $i$
    \item $B$ - terms where the rightmost operator is on site $i$, i.e. multisite terms that end on site $i$.
    \item $A$ - terms that neither start nor finish on site $i$, i.e. terms with diameter strictly greater than two\footnote{Diameter is the distance between the leftmost and rightmost non-trivial operator in a string.}.
\end{itemize}
Suppose the MPO for Hamiltonian $\hat{H}$ is in the form of Eq.~\eqref{supp-eq:MPO_standard_form}.
The MPO for $\hat{H}_D$ is given by
\begin{align}
    H^D &= \bra{v_L^D} W^{[0],D} W^{[1],D} \ldots W^{[N-1],D} \ket{v_R^D}, \\
    {W^{[i]}}^D_{\alpha \beta} &= \left[ \begin{array}{cccc} 
    \mathbb{I}^{4 \times 4} & C^{[i]} \otimes \mathbb{I}^{2 \times 2} & \mathbb{I}^{2 \times 2} \otimes \overline{C}^{[i]} & D^{[i]} \otimes \mathbb{I}^{2 \times 2} - \mathbb{I}^{2 \times 2} \otimes \overline{D}^{[i]} \\
    0 & A^{[i]} \otimes \mathbb{I}^{2 \times 2} & 0 & B^{[i]} \otimes \mathbb{I}^{2 \times 2} \\ 
    0 & 0 & \mathbb{I}^{2 \times 2} \otimes \overline{A}^{[i]} & -\mathbb{I}^{2 \times 2} \otimes \overline{B}^{[i]} \\ 
    0 & 0 & 0 & \mathbb{I}^{4 \times 4}
    \end{array}\right],
    \label{supp-eq:MPO_D_standard_form}
\end{align}
where the boundary vectors have been padded appropriately with zero entries.
If the original MPO had bond dimension $D$, the new MPO has bond dimension $2*D - 2$.

\textbf{Doubled space TDVP:} The doubled-space TDVP algorithm is unchanged operationally from the standard MPS algorithm when using MPO $H^D$ in place of MPO $H$, but some of its properties require reinterpretation,
In particular,  TDVP (at least in its original, symplectic formulation) preserves the expectation value of all conserved quantities that commute with the Hamiltonian $\hat{H}$, such as energy (always), magnetization, and other integrable conserved quantities.
In the doubled Hilbert space, TDVP will preserve properties that commute with $\hat{H}^D$---but these are no longer the physical conserved quantities..
In particular, the ``energy'' of $\hat{H}^D$ is conserved in time, but this is identically 0:
\begin{align}
    \langle \langle \hat{H}^D \rangle \rangle &= \langle \langle \rho(t) | \hat{H}^D | \rho(t) \rangle \rangle \nonumber \\
    &= \langle \langle \rho(0) | \hat{H}^D | \rho(0) \rangle \rangle \nonumber \\
    &= \mathrm{Tr} \bigg( \rho(0) \big[\hat{H} \rho(0) - \rho(0) \hat{H} \big] \bigg) = 0
\end{align}
Note that we have computed the expectation value in the doubled Hilbert space as one does for pure states, which is how we are treating the vectorized density matrix; i.e. we have both a bra and ket so effectively two copies of the density matrix rather than one as typically done for evaluating linear expectation values.
We must do this, since the TDVP algorithm works with respect to the vectorized density matrix and the inner product between vectorized states.
Hence, if an operator $\hat{O}$ commutes with $\hat{H}$, then $\hat{O} \otimes \mathbb{I}^{2^N \times 2^N}$ will commute with $\hat{H}^D$ and $\langle [\hat{O} \otimes \mathbb{I}^{2^N \times 2^N}](t) \rangle = \langle \hat{O} \otimes \mathbb{I}^{2^N \times 2^N} \rangle = \mathrm{Tr} ( \rho \hat{O} \rho)$.
So while this property is preserved by symplectic time evolution, it is not one we naturally think about and is not immediately relevant to charge transport given by $\mathrm{Tr} ( \rho \hat{O})$.

The single-site MPS formulation of TDVP is symplectic and thus preserves conserved quantities.
However, this algorithm does not grow the bond dimension during evolution.
So if starting from a product state MPS, as we do in the case of the spin domain wall density matrix, the bond dimension will remain trivial.
Instead, we use the two-site formulation of TDVP which allows the bond dimension to grow.
This algorithm is no longer symplectic so operators commuting with $\hat{H}^D$ are only conserved up to truncation error.
We refer the interested reader to a recent review for details of the single- and two-site TDVP algorithms as well as other alternative approaches to MPS time evolution~\cite{Paeckel_2019}.
Here, we typically evolve with an increasing timestep of $0.04-0.4-4.0$ for $50-45-X$ steps, where $X$ is determined by the desired final time.
This allows us to grow the bond dimension of the MPS with small step size so that the projection error of TDVP will be small before switching over to larger step sizes once a sizable basis of Schmidt states has been developed.
For anisotropic models with large $\Delta$ such that the operator norm of the Hamiltonian is large, we find time evolution to be more numerically stable if only steps of size $0.04-0.4$ are used.
For isotropic models, our results are insensitive to final step size, as we demonstrate in Fig.~\ref{supp-fig:hyperparameters}.

\subsection{Charge conservation in the doubled Hilbert space}
\label{supp-sec:charge_conservation}
Suppose that a Hamiltonian $\hat{H}$ commutes with charge operator $\hat{Q}$ and that initial state $\ket{\psi}$ is an eigenstate of $\hat{Q}$ with charge $Q$, $\hat{Q} \ket{\psi} = Q \ket{\psi}$.
Commonly one then places local constraints on the tensor entries in the MPS and MPO to constrain the state to be in the desired charge sector while also speeding up calculations by reducing the effective size of tensors.
We refer the interested reader to standard reviews and references therein~\cite{PhysRevB.83.115125,singh2012tensornetworkstatesalgorithms,Hauschild_2018,Schmoll_2020,Weichselbaum_2024}.

\begin{figure}
    \centering
    \includegraphics{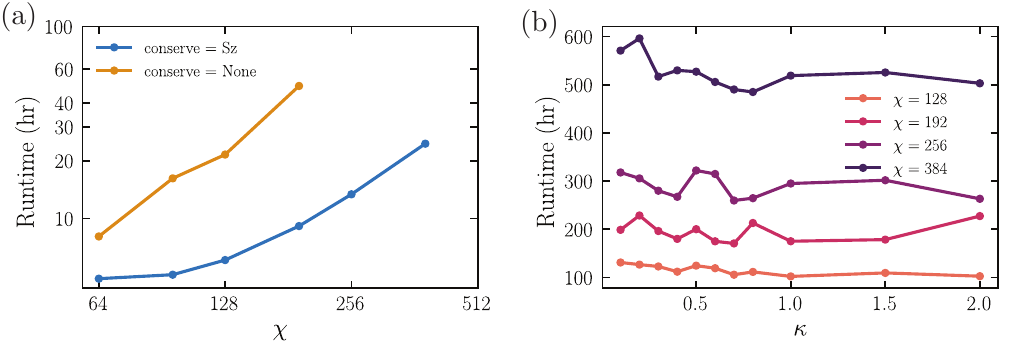}
    \caption{(a) Runtime as a function of bond dimension $\chi$ for doubled U(1) charge conserving and non-charge conserving dynamics, demonstrating the significant speedup from conserving the weak Sz symmetry. Simulation is of $L=1000$ Haldane Shastry chain to time $Jt=120$. \textcolor{red}{Longer time data} (b) Runtime as a function of Inozemtsev $\kappa$ and bond dimension for $L=2000$ simulation up to time $Jt = 800$, showing that runtimes are independent of $\kappa$. Each simulation uses a Hamiltonian with a $4$ exponent approximation, so MPO bond dimension is the same for all models.
    }
    \label{supp-fig:conserve}
\end{figure}

Here we extend this technique to charge conservation in the doubled Hilbert space.
The non-abelian SU(2) symmetry is not preserved in the TeNPy codebase we use~\cite{Hauschild_2018,Hauschild_2024} and additionally is not a symmetry of the anisotropic models with $\Delta \neq 1$.
Thus we focus on the abelian U(1) symmetry.
Rather than conserve $\hat{Q} = \sum_i \hat{s}^z_i$, we instead conserve the doubled $\hat{Q}^D = \hat{Q} \otimes \mathbb{I}^{2^N} - \mathbb{I}^{2^N} \otimes \hat{Q}$, which for magnetization counts the difference in charge between ket and bra legs on each site.
For the spin domain wall, even though it is a incoherent superposition of computational basis states of different magnetization, the density matrix has a doubled charge of $Q^D = 0$ since it is purely diagonal.
As we see from Fig.~\ref{supp-fig:conserve}, utilizing this charge symmetry gives at least a $5$x speedup, seemingly growing with bond dimension, and is paramount in enabling large-bond dimension simulations to times $Jt \sim 1000$.

\subsection{Long-range Hamiltonian as sum of exponentials}
\label{supp-sec:exponentials}
To perform time evolution with TDVP as described above, we need a tensor network representation of the (single-copy) Hamiltonian as a matrix product operator (MPO).
Given an arbitrary coupling potential $f(r)$, we approximate it as the sum of $K$ exponentials,
\begin{align}
    f(r) \approx \sum_{i=1}^K c_i e^{-\lambda_i r},
    \label{supp-eq:exp}
\end{align}
where the coefficients $c_i$ and decay rates $\lambda_i$ are chosen to minimize the error $||f(r) - \sum_{i=1}^K c_i e^{-\lambda_i r}||_2$.
This is standard practice for tensor network numerics involving power law interacting Hamiltonians, as one cannot exactly encode any power law interaction in a finite bond dimension MPO for an arbitrary system size~\cite{Pirvu_2010,Crosswhite_2008}.
For our XXZ-like Hamiltonians, the bond dimension for a $K$-exponential approximation is $3K + 2$.

\begin{figure}
    \centering
    \includegraphics{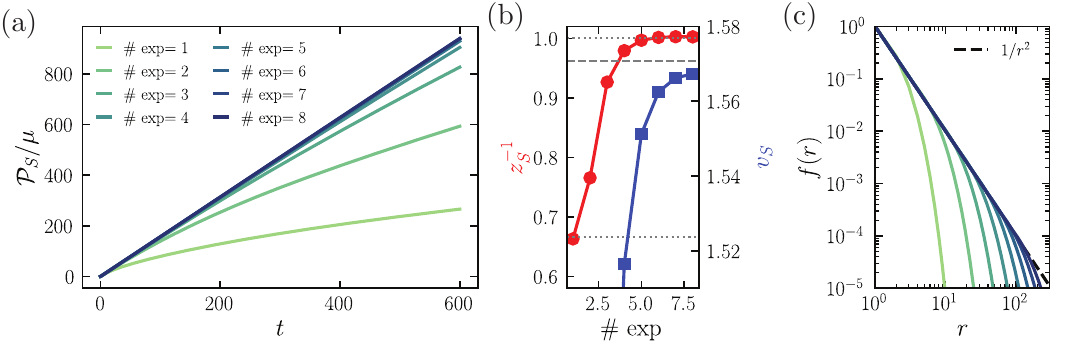}
    \caption{(a) Polarization transfer as a function of time, (b) inverse dynamical critical exponent at time $Jt=600$ (red) and spinon velocity (blue) extracted from slope of polarization transfer, and (c) interaction strength for spin transport of Haldane-Shastry Hamiltonian approximated by increasing number of exponents. In (b), the dotted line is $z^{-1}_S=1$, the expected dynamical critical exponent for ballisitc spin transport; the dashed line is $v_S = \pi/2 \approx 1.571$, the spinon velocity computed via generalized hydrodynamics~\cite{Bulchandani_2024}. Ballistic transport is observed at the latest times simulable for a chain of $N = 2000$ spins with $4$ exponents.
    }
    \label{supp-fig:exponents}
\end{figure}

By using exponentials to approximate a power-law $f(r) = \sfrac{1}{r^\alpha}$, we will by construction fail to capture the long-range tail.
See Fig.~\ref{supp-fig:exponents}(c) for approximations to a $f(r) = 1/r^2$ interaction with increasing number of exponents.
The model realized by $K$ exponentials can be viewed as a perturbation to the original power-law $f(r)$.
As an example, consider $T=\infty$ spin transport in the Haldane-Shastry model, $f(r) = \sfrac{1}{r^2}$.
We approximate $f(r)$ with between $K=1$ and $K=8$ exponentials and calculate the polarization transfer (Eq.~\eqref{supp-eq:PT}) and inverse dynamical critical exponent $z_S^{-1}(t)$ starting from a weak domain wall (Fig.~\ref{supp-fig:exponents}(a,b)).
While spin transport in the Haldane-Shastry model is expected to be ballistic, we instead see a crossover from ballistic $z_S^{-1}=1$ to sub-ballistic behavior as the number of exponents is decreased.
This crossover behavior is discussed through the lens of SU(2) symmetric perturbations to integrable models in Sec.~\ref{supp-sec:HS_spin} and Fig.~\ref{supp-fig:HS_perturbation}.
Also, note that the spinon velocity, extracted from the slope of the polarization transfer versus time (not the slope of log-log plot, which would give the instantaneous dynamical critical exponent $z_S^{-1}(t)$), is approaching $v_S = \pi/2$~\cite{Bulchandani_2024} with increasing exponents $K$.
For $K=8$ exponentials, we find $v_S = 1.567$.

Note that at large distances, the Inozemtsev potential is approximately the sum of 3 terms of the form of Eq.~\eqref{supp-eq:exp}:
\begin{align}
    \frac{\sinh^2(\kappa)}{\sinh^2(\kappa r)} 
    &\approx 4 \sinh^2(\kappa) \left( e^{-2 \kappa r} - e^{-6 \kappa r} + 2 e^{-4 \kappa r}\right).
\end{align}
\noindent Thus Inozemtsev models are naturally amenable to study with MPOs (see Fig.~\ref{supp-fig:KPZ_constant}(b) for Inozemtsev interaction potentials as a function of $\kappa$, showing exponential decay at large distances).

\begin{table}[]
\begin{tabular}{@{}l|llllllll@{}}
\toprule
\# exp       & 1      & 2      & 3      & 4      & 5       & 6      & 7      & 8      \\ \midrule
$\alpha=2$   & 0.3145 & 0.1051 & 0.047  & 0.0247 & 0.0143  & 0.0089 & 0.0057 & 0.0038 \\ \midrule
$\alpha=2.5$ & 0.1307 & 0.03   & 0.0099 & 0.0041 & 0.002   & 0.001  & 0.0006 & 0.0004 \\ \midrule
$\kappa=0.1$ & 0.2266 & 0.0391 & 0.0053 & 0.0005 & 3.3e-05 & 1e-06  & 0.0    & 0.0    \\ \midrule
$\kappa=0.4$ & 0.0714 & 0.0013 & 5e-06  & 0.0    & 0.0     & 0.0    & 0.0    & 0.0    \\ \bottomrule
\end{tabular}
\caption{Approximation error, measured by the L1 norm $\epsilon = \sum_r | f(r) - \sum_{i=1}^K c_i e^{-\lambda_i r}|$, for power laws with $\alpha=2, 2.5$ and Inozetmsev models with $\kappa=0.1, 0.4$ with varying number of exponents. Error is calculated up to distance $r=1000$, and errors less than $10^{-6}$ are reported as zero.}
\label{tab:approx_error}
\end{table}

In this work, we use $K=4$ exponents for all simulations.
In Table~\ref{tab:approx_error}, we show the L1-norm error from approxmating various power-law and Inozemtsev interactions with different number of exponents $K$.
The error for Inozemtsev models is very small for $K > 3$, due to the functional form at large distances discussed above.
As demonstrated in Fig.~\ref{supp-fig:exponents}, the finite number of exponentials will affect HS results on the timescales $Jt \sim 600$, which is beyond what was used for the structure factors and currents in the main text.
Power laws with $\alpha > 2$ and Inozemtsev models will be similarly affected by imperfections in the model, but we expect this to occur at times $Jt > 600$ as both the interaction range and approximation error with 4 exponentials for these models are smaller than that of the HS model..
Thus we do not expect model approximation error to be detectable in our results.

\subsection{Correctness of our method}
\label{supp-sec:correctness}
\begin{figure}
    \centering
    \includegraphics{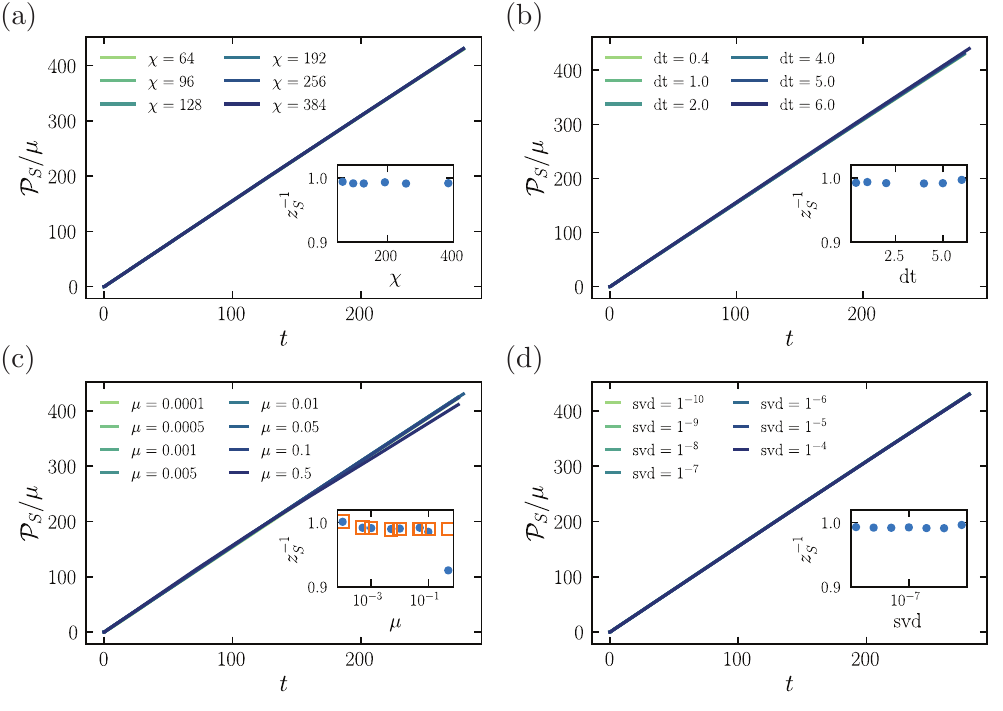}
    \caption{Polarization transfer (main figure) and inverse dynamical critical exponent (inset) of $T=\infty$ spin transport for Haldane-Shastry model with ballistic $z_S^{-1} = 1$ spin transport. Hyperparameters are (a) bond dimension $\chi$, (b) time step used in TDVP (after initial steps with smaller size), (c) domain wall strength $\mu$, and (d) cutoff of singular value decomposition (SVD) matrix factorization. The default parameters are $L=1000$, $\chi=128$, $\mathrm{d}t = 4.0$, $\mu=0.05$, and $\mathrm{svd}=10^{-6}$. In the inset of (c), the red boxes are for $\mu'$, which is the domain wall strength for the modified initial state $\tilde{\rho}$ which has no contributions beyond linear response.
    }
    \label{supp-fig:hyperparameters}
\end{figure}

Having discussed our numerical approach in detail, we now demonstrate that our results are largely robust to choices in initial state construction, TDVP evolution timestep, bond dimension, and SVD truncation precision.
In Fig.~\ref{supp-fig:hyperparameters} we consider spin transport in the Haldane-Shastry model and vary each of these parameters, where we find the expected ballistic behavior and consistent polarization transfer for a broad range of hyperparameters.
Interestingly, we see in the inset of Fig.~\ref{supp-fig:hyperparameters}(c) that deviation from ballistic spin transport is found with a spin domain wall of strength $\mu=0.5$, indicating that the beyond linear-response, higher-order terms present in the initial state can no longer be ignored.
However, the linear response initial condition with $\mu'=0.5$ (red boxes) shows ballistic transport, as one expects.

In this work, we will use a TDVP timestep of $\mathrm{d}t=4.0$ (after initial evolution steps with $\mathrm{d}t=0.04$ and $\mathrm{d}t=0.4$ to build up the bond dimension and the space into which the Hamiltonian is projected when evolving with TDVP), SVD truncation precision $10^{-6}$, domain wall strength $\mu=0.05$ and 4 exponents to approximate the Hamiltonian.
We typically use bond dimension $\chi=256$ or $\chi=384$.
We expect that the finite number of exponential approximation to the Hamiltonian, the finite domain wall strength, and the finite bond dimension each will lead to diffusive transport as $t \rightarrow \infty$, even for integrable models, due to numerical imperfections.
However, given our choice of parameters discussed above, we expect this to be far beyond the timescales investigated in this work and have not seen any evidence to the contrary.

\section{Charge Density and Current for Long-Range Models}
\label{supp-sec:current_formulas}
The current density in the evolved domain wall state is another signature of the dynamical universality class in addition to the dynamical critical exponent and two-point autocorrelation function.
In this section, we thus derive the form of the spin and energy current for 1D XXZ-type model with arbitrary potential $f(|i-j|)$ that depends only on the distance between two interacting sites. 
This will be applicable for the long-range XY, Haldane Shastry, and Inozemtsev models.

Consider a 1D lattice of quantum spin-1/2s. We will work with the general Hamiltonian in terms of \textit{Pauli} matrices
\begin{align}
    H &= \sum_{i<j} f_{|i-j|} (X_i X_j + Y_i Y_j + \Delta Z_i Z_j) \\
    &= \sum_{i<j} f_{|i-j|} \left(\frac{S^+_i S^-_j + S^-_i S^+_j}{2} + \Delta Z_i Z_j\right),
\end{align}
\noindent where we sum over both sites $i$ and $j$ but only count each interaction once. 
We do not put hats on operators in this section for notational convenience.
Define the spin raising and lowering operators as $S_i^+ = X_i + i Y_i$ and $S_i^- = (S_i^+)^\dagger = X_i - i Y_i$. 
Note that these are twice the usual spin raising and lowering operators since we have used Pauli matrices instead of spin-1/2 matrices:
\begin{align*}
    S_i^+ = \left( \begin{array}{cc} 0 & 2 \\ 0 & 0 \end{array} \right), \:\:\: S_i^- = \left( \begin{array}{cc} 0 & 0 \\ 2 & 0 \end{array} \right).
\end{align*}
Then $S^+_i S^-_j + S^-_i S^+_j = 2 (X_i X_j + Y_i Y_j)$. 
Since we want to use U(1) charge conservation in the tensor network numerics, we use $\{S^+, S^-, S^z\}$ and not $\{S^x, S^y, S^z\}$. 
The function $f(r)$ is model-dependent, and several choices are listed below.
\begin{align}
    f_r = \left\{ \begin{array}{ll} 
    \delta_{r,1} & \text{Nearest neighbor} \\
    \frac{1}{r^\alpha} & \text{power law} \\
    \frac{\left(\sinh{\kappa}\right)^2}{\left(\sinh{\kappa r}\right)^2} & \text{Inozemtsev} \\
    \end{array} \right.
\end{align}

\subsection{Spin current}
\label{supp-sec:spin_current}
The conserved charge is given by

\begin{align}
    Z^{T} = \sum_i Z_i,
\end{align}

\noindent where $Z_i$ is the conserved density on the \textit{site}. 
The current is defined in terms of $\dot{Z}_i$, the time derivative of the spin density.
This will not be sensitive to the anisotropy term $\Delta Z_i Z_j$.
\begin{align}
    \dot{Z}_i =& i [ H, Z_i] = j_x - j_{x+1} \nonumber \\ 
    =& i \sum_{a \neq i} f_{|a-i|} \left( S^+_a S_i^- - S^+_i S^-_a \right)
\end{align}
\noindent Note that we are using $\ket{\uparrow} = \ket{0} = \left( \begin{array}{c} 1 \\ 0 \end{array} \right)$ so that $S^+ \ket{\downarrow} = \ket{\uparrow}$. Then, $[S^-, Z_i] = 2 S^-$ and $[S^+, Z_i] = - 2 S^+$.

We want to identify the \textit{bond} spin current $j_x$ from this.
It is not immediately apparent how to define $j_x$ from $\dot{z}_x = j_x - j_{x+1}$ as one can arbitrarily shift $j_x$ and $j_{x+1}$ in opposite ways that would not effect the divergence.
So we use Eq. (3.10) of \cite{Pozsgay_2020} to remove this ambiguity.
\begin{align}
    j_x =& i \sum_{y=x}^{\infty} [ H,  Z_y] \nonumber \\
    =& i \sum_{y=x}^{\infty} \sum_{a < x} f_{|a-y|} \left( S^+_a S_y^- - S^+_y S^-_a \right)
\end{align}
Intuitively, this is the interaction of all sites strictly to the left of site $x$ with those strictly to the right of site $x-1$, or equivalently the terms that cross the bond from site $x-1$ to site $x$. 

In the case of a nearest-neighbor Hamiltonian with $f_{|a-b|} = \delta_{|a-b|, 1}$, we have the bond spin current
\begin{align}
    j_x &= i \left( S_{x-1}^+ S_x^- - S_x^+ S_{x-1}^- \right) \nonumber \\
    &= i \left( -2 i X_{x-1} Y_x + 2i Y_{x-1} X_x \right) \nonumber \\
    &= 2 X_{x-1} Y_x - 2 Y_{x-1} X_x \label{supp-eq:NN_spin_current}.
\end{align}

\subsection{Energy current}
\label{supp-sec:energy_current}
Unlike the spin current, the energy current is dependent on the anisotropy parameter $\Delta$. 
As before, we will derive the current from the time derivative of the energy density, so we define a site energy density which will yield a bond energy current density
\begin{align}
    h_{i} &= \frac{1}{2} \sum_{j \neq i} f_{|i-j|} \left(\frac{S^+_i S^-_j + S^-_i S^+_j}{2} + \Delta Z_i Z_j\right) \\
    &= \sum_{j \neq i} h_{i,j}.
\end{align}
This is half of the interactions between site $i$ and all other sites. 
The factor of $1/2$ is because we count each interaction between sites $i$ and $j$ twice, one in $h_i$ and once in $h_j$.

\subsubsection{Setup}
We now consider $[h_i, h_j]$. 
Since both of these are non-local, there are several cases to consider. 
We will first work out the commutator $[h_{a,b} , h_{c,d}]$ for generic indices. 
If no indices are common, then the commutator is zero.
It is not possible for 3 or 4 indices to be the same as $a \neq b$ and $c \neq d$. 
If $a=c$ and $b=d$ (or $a=d$ and $b=c$ since $h_{i,j} = h_{j,i}$), then the commutator is zero. 
So the only option to consider is (without loss of generality) $a=c$ and $b \neq d$, i.e. the two terms share one single index; all other cases can be found by relabeling.
\begin{align}
    [h_{a,b},h_{a,d}] =& [h_{b,a},h_{a,d}] = [h_{a,b},h_{d,a}] = [h_{b,a},h_{d,a}] \nonumber \\
    =& \frac{f_{|a-b|} f_{|a-d|}}{4} \left\{ S_b^- Z_a S_d^+ - S_b^+ Z_a S_d^- + \Delta \bigg( Z_b S_a^+ S_d^- - Z_b S_a^- S_d^+ - S_b^- S_a^+ Z_d + S_b^+ S_a^- Z_d \bigg) \right\}
    \label{supp-eq:e_comm}
\end{align}

To eventually use Eq. (3.10) of \cite{Pozsgay_2020}, we need $[h_{a\neq b}, h_b]$. 
To make this clear, consider an $L = 6$ system and set $a=2, b=4$. 
Then we can enumerate all the terms.
\begin{align*}
    [h_2, h_4] =& [h_{2,1} + h_{2,3} + h_{2,4} + h_{2,5} + h_{2,6}, \\
    & h_{4,1} + h_{4,2} + h_{4,3} + h_{4,5} + h_{4,6}] \\
    =& \sum_{i \neq 2, 4} [h_{2,i}, h_{4,i}] + [h_{2,4}, h_{4,i}] + [h_{2,i}, h_{4,2}]
\end{align*}
These $3(L-2) = 12$ terms can be encapsulated in the following formula.
\begin{align}
    [h_a, h_b] =& \sum_{c \neq a} \sum_{d \neq b} [h_{a,c}, h_{b,d}] \nonumber \\
    =& \underbrace{\sum_{d \neq b,a} [h_{a,d}, h_{b,d}]}_{c=d} + \underbrace{\sum_{d \neq b,a} [h_{a,b}, h_{b,d}]}_{c=b} + \underbrace{\sum_{d \neq a,b} [h_{a,d}, h_{b,a}]}_{d=a, c \rightarrow d}
    \label{supp-eq:e_comm_2}
\end{align}
Now, we want to find the commutator $[H, h_b] = \sum_{a \neq b} [h_a, h_b]$. In our six site example, consider $a = 2, i = 3$ and $a=3, i=2$.
\begin{align}
    [H, h_4] \ni& [h_{2,3}, h_{4,3}] + [h_{2,4}, h_{4,3}] + [h_{2,3}, h_{4,2}] \nonumber \\
    &+ [h_{3,2}, h_{4,2}] + [h_{3,4}, h_{4,2}] + [h_{3,2}, h_{4,3}] \nonumber \\
    \ni& 2 [h_{2,3}, h_{4,3}] + 2 [h_{2,3}, h_{4,2}] \nonumber 
\end{align}
Applying this intuition to the generic formula, consider the middle term of the generic formula. 
We see that this must be zero since for any pair $(a, d)$ such that $a \neq d$, $a \neq b$, and $d \neq b$, there will be another pair $(d, a)$. 
These will cancel due to $[A,B] = -[B, A]$ and $h_{i,j} = h_{j,i}$. So essentially, we are symmetrizing over $a$ and $d$. 
When we apply this to the middle term, we get 
\begin{align*}
    \sum_{a \neq b} \sum_{d \neq b,a} [h_{a,b}, h_{b,d}] &= \sum_{a \neq b} \sum_{d \neq b > a} [h_{a,b}, h_{b,d}] + [h_{d,b}, h_{b,a}] \\
    &= 0.
\end{align*}
Thus, we find the following expression
\begin{align}
    [H, h_b] &= \sum_{a \neq b} [h_a, h_b] \nonumber \\
    &= 2 \sum_{a \neq b} \sum_{d \neq b > a} [h_{a,d}, h_{b,d} + h_{b,a}].
\end{align}
Note that technically this allows us to define the current, as $\dot{h}_b = i[H, h_b] = j_b - j_{b+1}$. However, as in the case of the spin current, it is not clear how to define the individual current terms due to the additive ambiguity.

\subsubsection{Current definition}
\label{supp-sec:current_definition}
So finally we can use Eq.~(3.10) of \cite{Pozsgay_2020} and Eq.~\ref{supp-eq:e_comm} to unambiguously define the bond current.
\begin{align}
    j_x =& i \sum_{y=x}^{\infty} [ H, h_y] \nonumber \\
    =& i \sum_{y=x}^{\infty} \sum_{a < x} [h_a, h_y] \label{supp-eq:current}.
\end{align}
Note that the sum over $a$ can be restricted to terms to the left of site $x$ since terms with $a > x $ appear twice with opposite ordering in the commutator.
Consider the commutator given by Eq.~\ref{supp-eq:e_comm} and taking $a < x$ and $y \geq x$.
We now write this in a form such that the free index is always between $a$ and $y$. 
So we need to consider each term in this equation and break it up into cases: (1) $d < a$, (2) $a < d < y$ (what we want so nothing needs to be done), and (3) $y < d$.

\subsubsection{$\sum_{d \neq b,a} [h_{a,d}, h_{b,d}]$}
\noindent\textbf{Case I:} $d < a$. Relabel $(d,a,b) \rightarrow (a, d_<, b)$, where $d_<$ is a site to the left of site $x$. Then $[h_{a,d}, h_{b,d}] \rightarrow [h_{a,d_<}, h_{a,b}]$. This a type 3 term of Eq.~\ref{supp-eq:e_comm_2} with a $<$ restriction on $d$.

\noindent\textbf{Case II:} $a < d < b$. Nothing needs to be done; so this is a type 1 term of Eq.~\ref{supp-eq:e_comm_2} with no restriction on d.

\noindent\textbf{Case III:} $b < d$. Relabel $(a,b,d) \rightarrow (a, d_{\geq}, b)$, where $d_{\geq}$ is a site to the right of and including site $x$. Then $[h_{a,d}, h_{b,d}] \rightarrow [h_{a,b}, h_{b,d_{\geq}}]$. This a type 2 term of Eq.~\ref{supp-eq:e_comm_2} with a $\ge$ restriction on $d$.

\subsubsection{$\sum_{d \neq b,a} [h_{a,b}, h_{b,d}]$}
\noindent\textbf{Case I:} $d < a$. Relabel $(d,a,b) \rightarrow (a, d_<, b)$. Then $[h_{a,b}, h_{b,d}] \rightarrow [h_{d_<, b}, h_{a,b}] = -[h_{a,b}, h_{d_<, b}]$. This a type 2 term of Eq.~\ref{supp-eq:e_comm_2} with a $<$ restriction on $d$. Note the $-$ sign. This arises from changing the order of the commutator.

\noindent\textbf{Case II:} $a < d < b$. Nothing needs to be done; so this is a type 2 term of Eq.~\ref{supp-eq:e_comm_2} with no restriction on d.

\noindent\textbf{Case III:} $b < d$. Relabel $(a,b,d) \rightarrow (a, d_{\geq}, b)$. Then $[h_{a,b}, h_{b,d}] \rightarrow [h_{a,d_{\geq}}, h_{b,d_{\geq}}]$. This a type 1 term of Eq.~\ref{supp-eq:e_comm_2} with a $\geq$ restriction on $d$.

\subsubsection{$\sum_{d \neq a,b} [h_{a,d}, h_{b,a}]$}
\noindent\textbf{Case I:} $d < a$. Relabel $(d,a,b) \rightarrow (a, d_<, b)$. Then $[h_{a,d}, h_{b,a}] \rightarrow [h_{a, d_<}, h_{d_<,b}]$. This a type 1 term of Eq.~\ref{supp-eq:e_comm_2} with a $<$ restriction on $d$.

\noindent\textbf{Case II:} $a < d < b$. Nothing needs to be done; so this is a type 3 term of Eq.~\ref{supp-eq:e_comm_2} with no restriction on d.

\noindent\textbf{Case III:} $b < d$. Relabel $(a,b,d) \rightarrow (a, d_{\geq}, b)$. Then $[h_{a,d}, h_{a,b}] \rightarrow [h_{a,b}, h_{a,d_{\geq}}] = -[h_{a,d_{\geq}}, h_{a,b}]$. This a type 3 term of Eq.~\ref{supp-eq:e_comm_2} with a ${>=}$ restriction on $d$. Note the $-$ sign. This arises from changing the order of the commutator.

\subsubsection{Putting it together}
\label{supp-sec:together}
Let us now add these together so that we can rewrite Eq.~\ref{supp-eq:current} with $d$ in between $a$ and $b$.
\begin{align}
    j_x =& i \sum_{y=x}^{\infty} \sum_{a < x} [h_a, h_y] \nonumber \\
    =& 2i \sum_{a < x} \sum_{y=x}^{\infty} \bigg( \big(\sum_{a < d < x} + \sum_{x \leq d < y} \big) [h_{a,d}, h_{d,y}] + \sum_{x \leq d < y} [h_{a,y}, h_{d,y}] + \sum_{a < d < x} [h_{a,d}, h_{a,y}] \bigg) \nonumber \\
\end{align}  
We now need to use Eq.~\ref{supp-eq:e_comm} to simply the commutators and combine terms where possible. 
Let us first consider the sum $\sum_{a < d < x}$.
\begin{align}
    \sum_{a < d < x} \supset& \: [h_{a,d}, h_{d,y}] + [h_{a,d}, h_{a,y}] \nonumber \\
    =& \frac{f_{|a-d|}}{4} \left\{ \bigg( S_a^- Z_d S_y^+ ( f_{|d-y|} - \Delta f_{|a-y|}) + S_a^+ Z_d S_y^- ( \Delta f_{|a-y|} - f_{|d-y|}) \bigg) \right. \nonumber \\
    &+ \left. \bigg( Z_a S_d^+ S_y^- ( \Delta f_{|d-y|} - f_{|a-y|}) + Z_a S_d^- S_y^+ ( f_{|a-y|} - \Delta f_{|d-y|}) \bigg) \right. \nonumber \\
    &+ \left. \bigg( \Delta S_a^+ S_d^- Z_y (f_{|d-y|} - f_{|a-y|}) + \Delta S_a^- S_d^+ Z_y (f_{|a-y|} - f_{|d-y|}) \bigg) \right\} \label{supp-eq:curr_term1}
\end{align}  
Now consider the sum $\sum_{x \leq d < y}$.
\begin{align}
    \sum_{x \leq d < y} \supset& \: [h_{a,d}, h_{d,y}] + [h_{a,y}, h_{d,y}] \nonumber \\
    =& \frac{f_{|d-y|}}{4} \left\{ \bigg( S_a^- Z_d S_y^+ ( f_{|a-d|} - \Delta f_{|a-y|}) + S_a^+ Z_d S_y^- ( \Delta f_{|a-y|} - f_{|a-d|}) \bigg) \right. \nonumber \\
    &+ \left. \bigg( \Delta Z_a S_d^+ S_y^- ( f_{|a-d|} - f_{|a-y|}) + \Delta Z_a S_d^- S_y^+ ( f_{|a-y|} - f_{|a-d|}) \bigg) \right. \nonumber \\
    &+ \left. \bigg( S_a^+ S_d^- Z_y (\Delta f_{|a-d|} - f_{|a-y|}) + S_a^- S_d^+ Z_y (f_{|a-y|} - \Delta f_{|a-d|}) \bigg) \right\} \label{supp-eq:curr_term2}
\end{align}  

So these last two expressions tell us to how calculate $j_x$ as a linear combination of 3-body operators.

\subsubsection{Measuring current in TNS calculations}
\label{supp-sec:measuring}
The bond energy current is not easy to measure in TNS calculations. 
While in principle we could construct an MPO for all the terms for a single bond current $j_x$, this is non-trivial due to the $f_{r}$ factors.
If we use $K$ exponentials for $f$, then we would need $K^2$ exponentials for $f * f$. 
Instead, we can measure the terms $S_a^+ Z_d S_y^-$, $Z_a S_d^+ S_y^-$, and $S_a^+ S_d^- Z_y$ for all combinations $a < d < y$ from $1, \ldots, L$. 
The Hermitian conjugate of these operators gives us the other 3 types of terms, so the complex conjugate of the expectation value will be sufficient to calculate the bond current. 
However, there will be order $\mathcal{O}(L^3)$ terms to measure, so we find the measurement of the bond current to be very computationally expensive.
In practice, we impose an interaction cutoff and do not consider sites separated by distance $r$ where $f(r) < 10^{-10}$.
The reduces the number of measures, yet still the calculation takes multiple days, given the large system sizes.

\subsection{Correctness of current equations}
\label{supp-sec:current_correctness}
\begin{figure}
    \centering
    \includegraphics{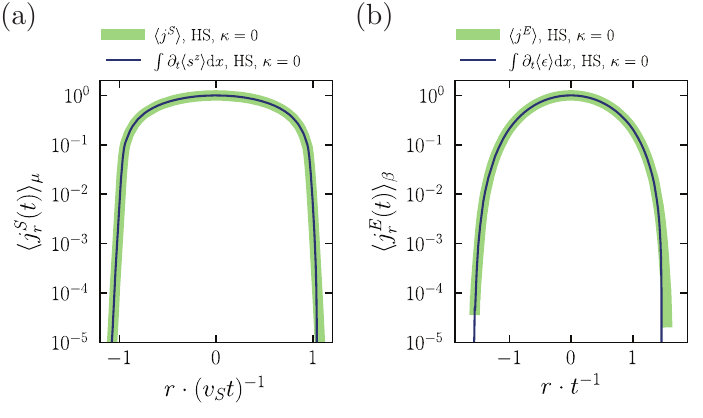}
    \caption{Comparison between explicit formula for current and current derived from the conserved charge density for (a) spin and (b) energy, showing excellent agreement. For both figures, the model is Haldane-Shastry, and data is measured at $Jt=440$ and $Jt=320$ for spin and energy, respectively.
    }
    \label{supp-fig:correctness}
\end{figure}

The correctness of our spin and energy current expressions can be ascertained by comparing the explicit expressions to the current found by integrating the time derivative of the charge density.
\begin{align}
    \dot{q}_x = j_x - j_{x+1} \rightarrow \sum_{y \leq x} \dot{q}_y = j_{x+1},
\end{align}
which is equivalent to how we unambiguously define the current.
In Fig.~\ref{supp-fig:correctness}, we compare the spin and energy current density from the two methods for the Haldane-Shastry model after evolving a weak-domain wall state.
We find that the two methods agree except for small differences at sites $r \gg 0$, where the integrated derivative of the charge density is expected to be sensitive to numerical error from all previous sites.
We thus explicitly measure the current density using the derived formulas.

\section{Integrable Models -- Spin: Additional Results}
\label{supp-sec:integrable_spin}
Here we present additional results for the infinite temperature spin transport for the Haldane-Shastry model, including comparisons of two-point autocorrelators and current densities to generlized hydrodynamics, timescale for deviation from ballistic transport in the perturbed Haldane-Shastry model, and KPZ-like transport for Inozemtsev models with various $\kappa$.

\subsection{Haldane-Shastry}
\label{supp-sec:HS_spin}
\begin{figure}
    \centering
    \includegraphics{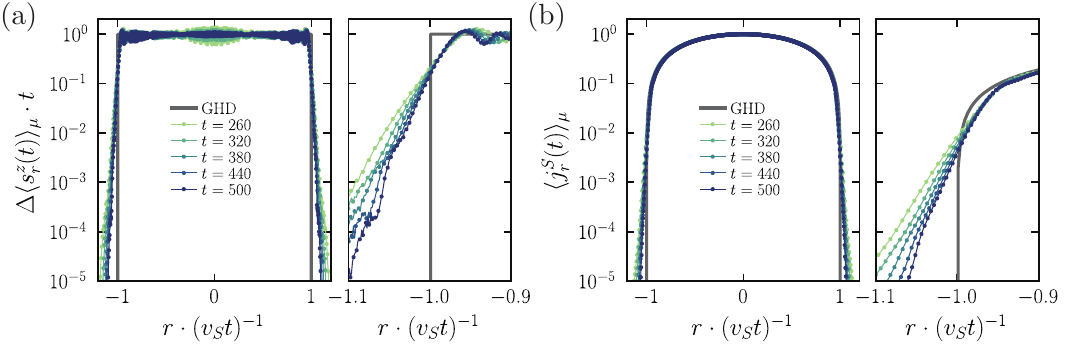}
    \caption{(a) Spin autocorrelation function and (b) spin current for the Haldane-Shastry model, compared to generalized hydrodynamics calculations~\cite{Bulchandani_2024}. Data at several times is shown and is rescaled by time and spinon velocity $\pi/2$, indicating a collapse with dynamical critical exponent $z_S = 1$. The right part of each figure zooms into data around $r \dot (v_S t)^{-1} = 1$, showing the increasing agreement with GHD predictions as time increases.
    }
    \label{supp-fig:HS_spin}
\end{figure}
\textbf{Comparison to generalized hydrodynamics:} Spin transport in the Haldane-Shastry model is known to be ballistic due to the conserved spin current~\cite{Sirker_2011}.
In the main text, we showed that linear profile of the melting of the spin domain wall, which collapses when rescaled by $t$; this confirms the ballistic transport with dynamical critical exponent $z_S = 1$.
Recently, the enhanced symmetry of the Haldane-Shastry model was used in a generalized hydrodynamics (GHD) framework to compute the spin two-point autocorrelation function and current density at infinite temperature~\cite{Bulchandani_2024}.
In Fig.~\ref{supp-fig:HS_spin} we compare our numerical measurements to these GHD predictions of two-point (a) and current (b), finding good agreement.
The box nature of the two-point function is better captured as time increases.
This is reminiscent of approximating a box potential with a finite sum of Fourier components; increasing the number of components (i.e. time) leads to a sharper edge of the box potential.
We take this as further confirmation, in addition to the benchmarking in Fig.~\ref{supp-fig:hyperparameters}, that our numerical approach produces accurate hydrodynamic results at late times.

\textbf{Transport in a perturbed integrable model:}
\begin{figure}
    \centering
    \includegraphics{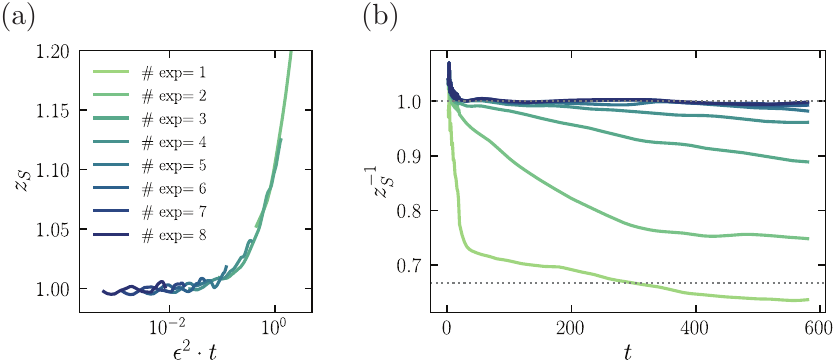}
    \caption{(a) Onset of sub-ballistic behavior in perturbed Haldane-Shastry models, i.e. models with the $1 / r^2$ interaction approximated by a finite number of Hamiltonians. The data collapses with $\epsilon^2 t$ where $\epsilon$ is the error in model approximation, indicating that the onset of sub-ballistic behavior occurs on a timescale $t^* \sim 1/\epsilon^2$. (b) Dynamical critical exponent as a function of time for various models, showing finite time flow.
    }
    \label{supp-fig:HS_perturbation}
\end{figure}
Consider a Hamiltonian $H = H_{\text{int.}} + \epsilon H'$ where $H_{\text{int.}}$ is an integrable model with a non-Abelian symmetry and $H'$ is a generic integrability-breaking perturbation with strength $\epsilon$.
While spin transport in $H_{\text{int.}}$ is conjectured to be anomalous, as evidenced by the ballistic and superdiffusive transport found in Haldane-Shastry and Inozemtsev models, respectively, transport with $H'$ is expected to be diffusive as $t \rightarrow \infty$.
However, the timescale at which this occurs is generally unknown, and several recent studies have numerically investigated the crossover to diffusion~\cite{McCarthy_2024,McRoberts_2024,Wang_2025}, finding a timescale $t^*\sim\epsilon^\alpha$ for power law $\alpha \in [3, 8]$.
Two of these works, Refs.~\cite{McCarthy_2024,Wang_2025} conjecture from both classical and quantum numerics that if $H'$ respects the non-Abelian symmetry of $H_{\text{int.}}$, then the crossover to diffusion occurs at a timescale $t^* \sim 1 / \epsilon^6$.

Here we use the finite exponential approximations to the Haldane-Shastry model to investigate the crossover to diffusion when starting with ballistic spin transport.
The models with $K$ exponents (Fig.~\ref{supp-fig:exponents}(c)) are expected to be non-integrable as the interaction is not fine-tuned.
However, the models have SU(2) symmetry since they are still Heisenberg-like and isotropic.
In Fig.~\ref{supp-fig:HS_perturbation}(b), we show the dynamical critical exponent as a function of time, as extracted from the slope of the polarization transfer (Fig.~\ref{supp-fig:exponents}(a)).
We see that $z_S^{-1}$ for the model with 1 exponent crosses below $2/3$ and presumably will approach a value of 1/2 (indicative of diffusion) as $t \rightarrow \infty$.
The models with finite $K$ can be viewed as perturbations to the true Haldane-Shastry model using Eq.~\eqref{supp-eq:exp}.
The perturbation strength is given by
\begin{align}
    \epsilon_K = \sum_r \left| \frac{1}{r^2} - \sum_{i=1}^K c_i e^{-i \lambda_i r} \right|,
\end{align}
which is the $L_1$-norm of the difference between interaction potentials.
We find that $z_S(t)$ collapses nicely with $\epsilon^2 t$, indicating that the timescale for the onset of diffusion is $t^* \sim \epsilon^2$.
The discrepancy between $\alpha=2$ found here for the Haldane-Shastry model and $\alpha=6$ suggested in classical and quantum numerics of short-range models is another sign of the peculiarity of the Haldane-Shastry model, in addition to the ballistic instead of superdiffusive spin transport for integrable models with a non-Abelian symmetry.
Note, however, that these approximate Haldane-Shastry models with $K$ exponents can \textit{also} be viewed as SU(2) symmetric perturbations to Inozemtsev models, leading to a slow crossover from KPZ-like spin superdiffusion to diffusion.
Hence, an arbitrary isotropic Heisenberg-like model represented as a sum of exponentials can be viewed as a perturbation to Haldane-Shastry with strength $\epsilon_{\mathrm{HS}}$ or a perturbation to the closest Inozemtsev model with strength $\epsilon_{\mathrm{Inoz.}}$.
The observed crossovers in transport will depend both on the crossover exponent $\alpha$ and the magnitude of the perturbations $\epsilon$.

We conjecture that Inozemtsev models, if appropriately perturbed, would show a crossover to diffusion on a timescale $t^* \sim \epsilon^6$, as other features of spin transport are consistent with that of the NN model.
In Table~\ref{tab:alpha_to_kappa}, we show the closest Inozemtsev model $f^{(\kappa)}(r) = \sinh(\kappa)^2 / \sinh(\kappa r)^2$ for several different power-law-interacting models $f^{(\alpha)}(r) = 1/r^\alpha$, treating $\kappa$ as a variational parameter and minimizing the $L_1$ difference $\epsilon = \sum_r | f^{(\alpha)}(r) - f^{(\kappa)}(r)|$.
The Inozetmsev model itself is then approximated by a sum of exponentials.
However, the approximation error to a true Inozemtsev interaction with $K$ exponents is small due to the exponentially decaying tail in the true interaction.
So, if one wanted to study crossover from superdiffusion to diffusion by perturbing Inozetmsev models, different perturbation than finite exponentials, as was used for the Haldane-Shastry model, is needed, such as a sinusoidal envelope leading to a non-monotonically-decreasing interaction.

\begin{table}[]
\begin{tabular}{@{}l|lllllll@{}}
\toprule
\textbf{$\alpha$}   & 2.0 & 2.2    & 2.5    & 3.0    & 4.5    & 6.0    & $\infty$ \\ \midrule
\textbf{$\kappa$}   & 0.0 & 0.179  & 0.469  & 0.881  & 1.512  & 2.063  & $\infty$ \\ \midrule
\textbf{$\epsilon$} & 0.0 & 0.0945 & 0.0881 & 0.0525 & 0.0083 & 0.0015 & 0.0 \\ \bottomrule
\end{tabular}
\caption{Viewing power-law interacting models as weak, SU(2) preserving perturbations of the closest Inozemtsev model by treating $\kappa$ as a variational parameter and minimizing $\epsilon = \sum_r |f^{(\alpha)}(r) - f^{(\kappa)}(r)|$.}
\label{tab:alpha_to_kappa}
\end{table}

\subsection{Inozemtsev}
\label{supp-sec:Inozemtsev_spin}
\begin{figure}
    \centering
    \includegraphics[width=0.95\textwidth]{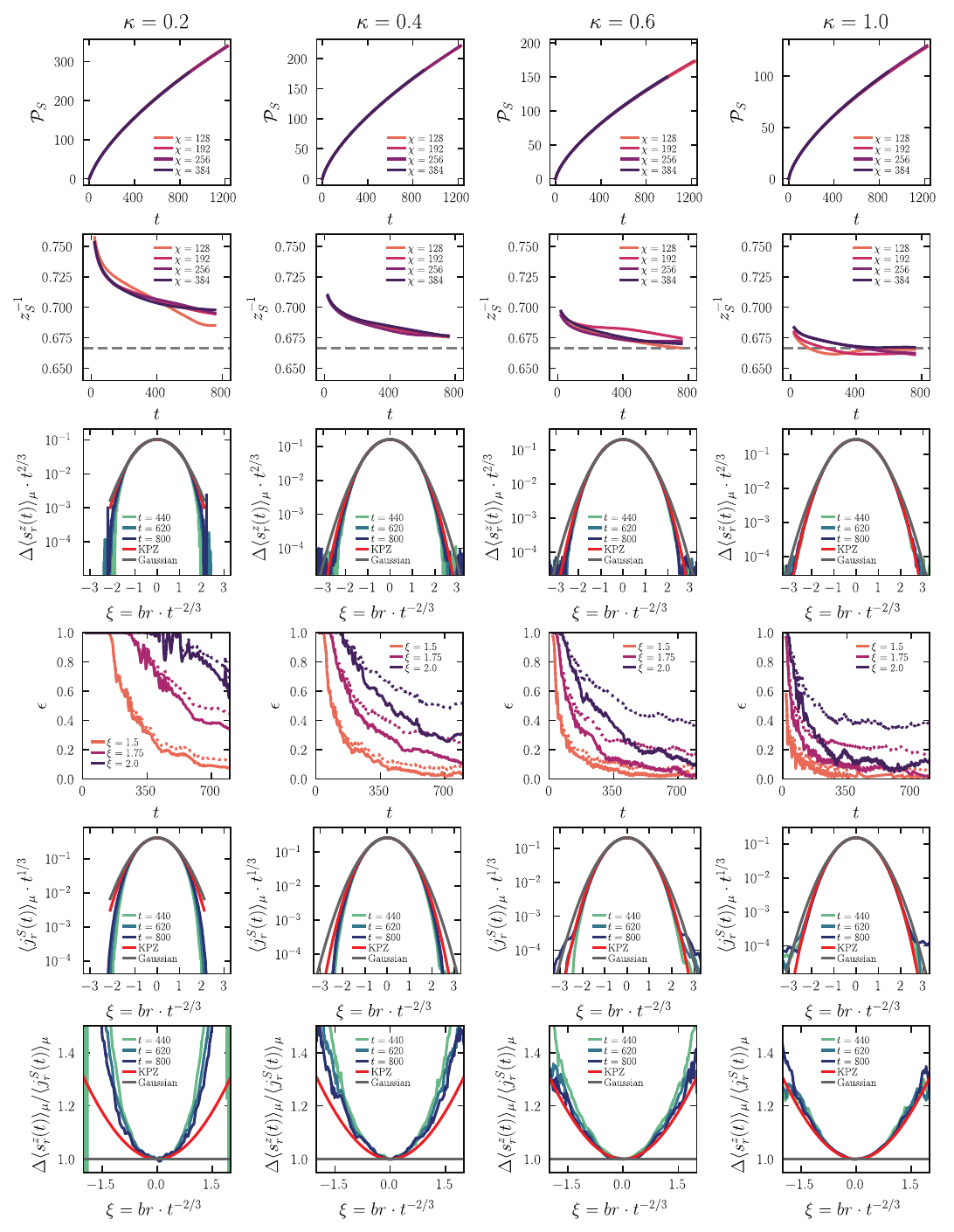}
    \caption{Spin transport for Inozemtsev models with various kappa. Columns correspond to $\kappa = 0.2, 0.4, 0.6, 1.0$, respectively. Rows are as follows: polarization transfer for various bond dimensions; dynamical critical exponent extracted from slope of polarization transfer; two-point spin autocorrelation function collapsed in time; error in spin two-point function when compared to KPZ (solid) and Gaussian (dotted) scaling predictions; spin current density collapsed in time; ratio of spin two point and current. Data for bottom four rows is for $\chi=384$.
    }
    \label{supp-fig:Inoz_spin}
\end{figure}
\begin{figure}
    \centering
    \includegraphics{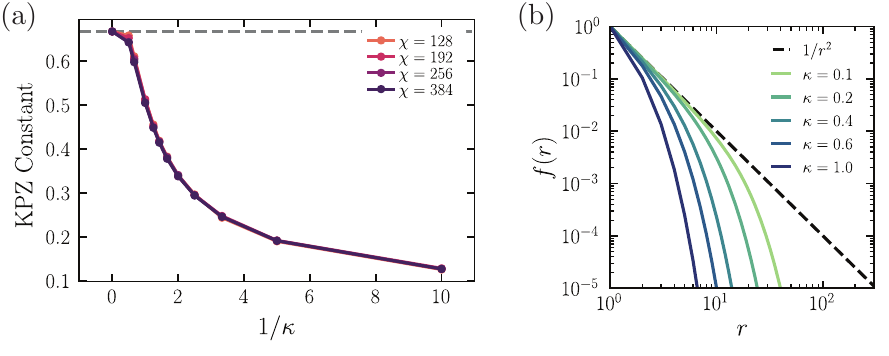}
    \caption{(a) Model dependent parameter $b$ found when fitting spin two-point or current density to KPZ predictions. (b) Interaction for various $\kappa$, demonstrating the exponential decay at large distances.}
    \label{supp-fig:KPZ_constant}
\end{figure}

\textbf{Polarization transfer and dynamical critical exponent:} We begin by showing in Fig.~\ref{supp-fig:Inoz_spin} additional spin transport data for $\kappa=0.2, 0.4, 0.6, 1.0$, with potentials given in Fig.~\ref{supp-fig:KPZ_constant}(b).
First, the polarization transfer as a function of bond dimension shows excellent convergence up to times $Jt \sim 1200$.
The dynamical critical exponent, found from the instantaneous slope of the polarization transfer, again shows excellent convergence in bond dimension.
While the larger $\kappa$ models have reached $z_S^{-1}=2/3$, the smaller models still show faster transport; these models can be viewed as integrable perturbations to the Haldane-Shastry model, so crossover from ballistic spin transport to the KPZ-like superdiffusion seemingly occurs at later times than $Jt \sim 800$.

\textbf{Spin two-point function and KPZ constant:} Next we look at the spin two-point autocorrelation function, rescaled by $t^{-2/3}$.
We compare to the best fit Gaussian and KPZ scaling function, where the later is obtained by an exact solution to a model in the KPZ universality class.
For the latter, there is a model-dependent fit parameter $b$ known as the KPZ constant, which we show in Fig.~\ref{supp-fig:KPZ_constant}(a) for various $\kappa$.
This constant for the nearest-neighbor model with $\kappa=\infty$ was conjectured, based on numerical simulations to be $b(\kappa=\infty) = 2/3$~\cite{Ljubotina_2019} and from generalized hydrodynamics calculations~\cite{De_Nardis_2020} to be $0.646$.
We find that as $\kappa \rightarrow \infty$, the fit of the spin two-point function for the Inozemtsev models yield a KPZ constant consistent with both predictions up to numerical accuracy.

\textbf{Error, current and ratio:} To show that the two-point scaling function is in better agreement with the KPZ form rather than the Gaussian form, we computer the relative error $\epsilon$ between numerical data and scaling predictions $f(\xi)$ at fixed $\xi = b r \cdot t^{-2/3}$ as a function of time:
\begin{align}
    \epsilon = \frac{\bigg|[\Delta \langle s^z_r(t) \rangle_\mu \cdot t^{2/3}](\xi) - f(\xi) \bigg|}{f(\xi)}.
\end{align}
For each $\kappa$ and $\xi$, the error with respect to the KPZ prediction is decreasing with time and is smaller than that with respect to the Gaussian prediction.
The current density again is more consistent with the KPZ prediction at late times than with the Gaussian function.
Finally, we consider the ratio of spin two-point to spin current, which allows us to rule out rescaled diffusion as the mechanism for superdiffusive transport~\cite{Ljubotina_2019}.
While the larger $\kappa$ are clearly consistent with the KPZ predictions, the smaller $\kappa$ are still approaching the KPZ prediction in time, despite simulating to $Jt \sim 800$.
Thus, while we conjecture that spin transport for all finite $\kappa$ Inozemtsev models falls into the same dynamical universality class as that of the nearest-neighbor Heisenberg model, further study is needed.

\section{Disordered Power Laws}
\label{supp-sec:disorder}
\begin{figure}
    \centering
    \includegraphics{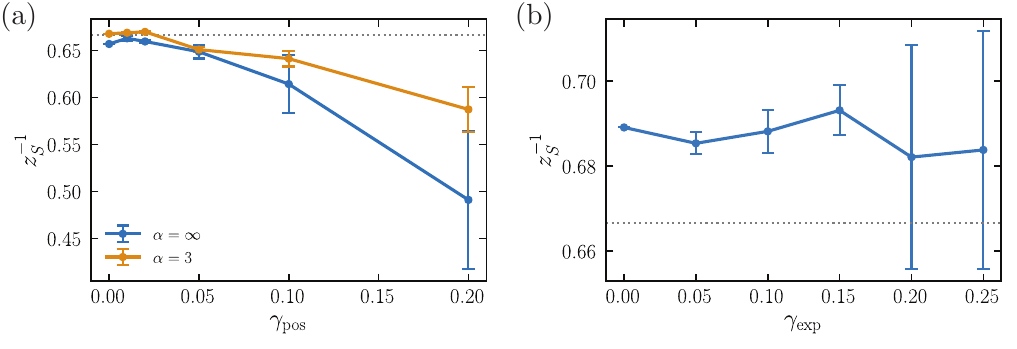}
    \caption{Inverse dynamical critical exponent for disordered models, averaged over multiple different disorder realizations, at time $Jt = 480$. Error bars are the standard deviation. (a) Spin-1/2 particle $i$ is placed at position $x_i + \gamma_{\text{pos}} \eta_i$, where $\delta$ is the disorder strength and $\eta_i$ is a random variable, $\eta_i \in [-0.5, 0.5]$. (b) Interaction is ``disordered" by $f(r_i) \rightarrow f(r_i + \gamma_{\text{exp}} \cos(\eta r_i))$ with $\eta$ an integer.}
    \label{supp-fig:disorder}
\end{figure}
In the main text we show that transport in power law models shows signs of integrability, despite the models being non-integrable.
Here we investigate whether positional disorder is enough to restore the expected diffusive spin transport.
We model disorder in two different ways.
First, we randomly disorder the locations of the position $x_i$ of the spin-1/2 particle $i$ with strength $\gamma_{\text{pos}}$:
\begin{align}
    x_i \rightarrow x_i + \gamma_{\text{pos}} \eta_i,
\end{align}
where $\eta_i$ is a random variable drawn from $[-0.5,0.5]$.
In Fig.~\ref{supp-fig:disorder}(a), we show the resulting spin dynamical critical exponent at $Jt=140$ for $\alpha=3$ and the nearest neighbor $\alpha=\infty$ models, averaging over 5 different disorder realizations.
We find that for small disorder strength $\gamma_{\text{pos}}$, the transport is not affected, as one intuitively expects.
The nearest-neighbor model is more susceptible to increased disorder strength, as the spin transport becomes diffusive with larger error bars; spin transport in the $\alpha=3$ model is still superdiffusive (again expected to be a finite time effect with diffusion as $t \rightarrow \infty$) with $3/2 < z_S < 2$ and comparatively smaller error bars.
Thus we conjecture that this anomalous behavior can still be observed in experiment, even with sizable disorder.

Note that the interaction becomes site-dependent since we have directly disordered the location of the spins.
Thus we must add each spin pair coupling term to the Hamiltonian MPO directly to the MPO, which leads to large MPO bond dimension.
In practice for the $\alpha=3$ model, we cutoff all interactions with strength $f(|x_i - x_j|) < 10^{-2}$.
In Fig.~\ref{supp-fig:disorder} we present an alternative approach for directly perturbing the long-range Hamiltonian in a way that preserves translational-invariance and effecnt representation as an MPO.
Rather than approximate $f(r) = 1/r^3$ as a sum of exponentials, we use the interaction $f(r) \rightarrow f(r + \gamma_{\text{exp}} \cos(\eta r))$ for integer $\eta$. 
We average over results from $\eta = 1, 2, 4, 5$ and find that the superdiffusive transport is stable, albeit with larger error bars, as ``disorder" strength $\delta$ is increased.
Again, this further supports our conjecture that this anomalous transport is stable to perturbations and can be observed in experiment.

\section{Integrable Models -- Energy: Additional Results}
\label{supp-sec:integrable_energy}
\begin{figure}
    \centering
    \includegraphics{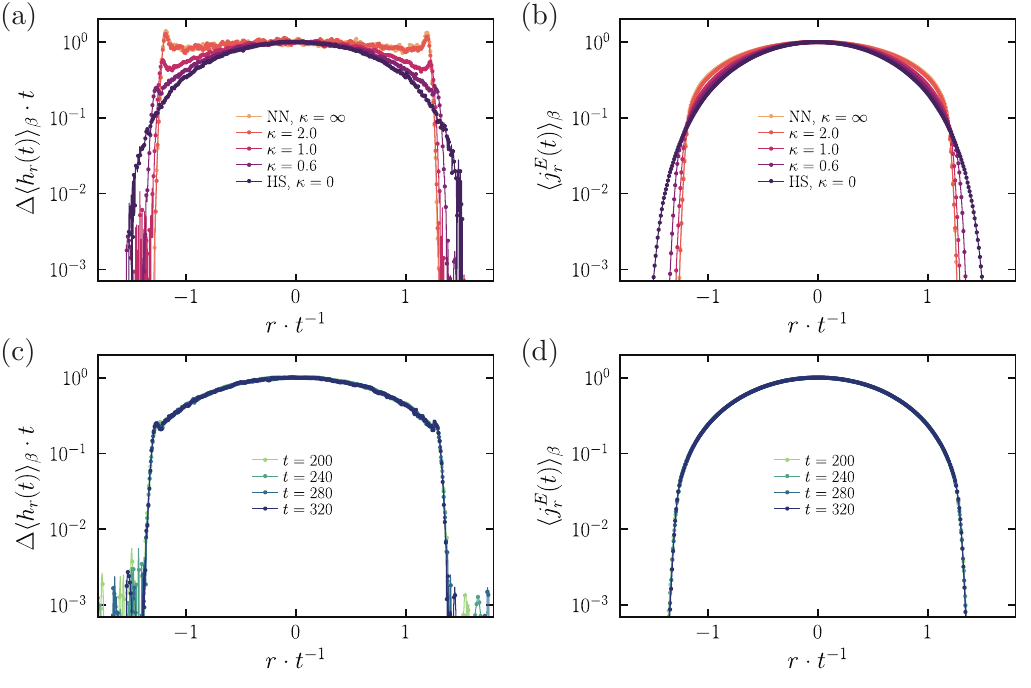}
    \caption{Energy two point, autocorrelation function (a) and current density (b) for various $\kappa$, including the Haldane-Shastry and nearest-neighbor Heisenberg model, rescaled by time. Energy two point (c) and current (d) show excellent collapse in time, shown for the $\kappa=0.6$ Inozemtsev model,}
    \label{supp-fig:energy}
\end{figure}

As we showed in the main text, energy transport across the range of $\kappa$, from Haldane-Shastry ($\kappa=0$) to nearest-neighbor Heisenberg ($\kappa=\infty$), is ballistic, as one expects for integrable models.
However, the profile of the melting of the energy domain wall, when rescaled by time, does not collapse to the same profile.
We find that the energy two-point autocorrelation function (Fig.~\ref{supp-fig:energy}(a)) and current density (Fig.~\ref{supp-fig:energy}(b)) show ballistic behavior yet again have visible differences.
Notably, the energy two-point function develops cusps and a box-signal-like shape as $\kappa \rightarrow \infty$.
The energy current is itself a conserved quantity for the nearest-neighbor Heisenberg model, which leads to the linear energy density profile from domain-wall melting; this is reminiscent of the linear spin density profile in the Haldane-Shastry model where the spin current is completely conserved.
Hence, we conjecture that for $\kappa < \infty$, the energy current is not a conserved quantity itself and only has overlap with a conserved quantity; this still yields ballistic energy transport but an energy density profile from domain-wall melting with non-linearities.

\section{Spin transport in a charged background and Drude weight}
\label{supp-sec:Drude}
Thus far we have studied spin transport at infinite temperature and in the net-zero magnetization sector; the initial equilibrium state is an equal weight superposition of all computational basis states so $\sum_i \hat{s}_i^z = 0$ on average.
It is known that spin transport in the Heisenberg model in a magnetically charged background is ballistic and thus hosts a non-zero Drude weight~\cite{De_Nardis_2020}.
Here we investigate whether the similarities between spin transport in the Heisenberg and Inozsemtsev models extend from a neutral to a charged background, i.e. do Inozsemtsev models support ballistic spin transport with a finite magnetization ensemble.
We find that at any finite magnetization $\delta$, spin transport in the Inozemtsev models also becomes ballistic.
Additionally, we then calculate the Drude weight as a function of both charge $\delta$ and Inozemtsev $\kappa$ and compare to generalized hydrodynamics results where available.
We find that the Inozemtsev Drude weights smoothly interpolate between those of the Haldane-Shastry and nearest-neighbor Heisenberg model.

First, we describe how to study transport in a charged background.
We modify the spin domain wall of strength $\mu$, given by Eq.~\ref{supp-eq:spin_DW}, to
\begin{align}
    \rho_i = \frac{\mathbb{I}_i^{2 \times 2}}{2} + \delta \hat{s}_i^z + \mu \hat{s}_i^z,
    \label{supp-eq:charged_spin_DW},
\end{align}
on the left half; for the right half, the third term has a minus sign due to the negative potential $e^{-\lambda \hat{s}_i^z}$; again, note that $\mu = \tanh\left(\frac{\mu}{2}\right)$.
For $\lambda = 0$, the average expectation value $\hat{s}^z$ is $\delta / 2$, so $\delta = 1$ yields the all up state.
Imposing a weak domain wall with $\lambda \ll 1$ allows us to study the linear response relaxation back to equilibrium.
Note that this initial state again has bond dimension 1 and the weak Sz symmetry.
We use $\mu = 0.005$.

We calculate the Drude weight from the \textit{connected} two-point spin auto-correlation function $C(x,t) = \langle \hat{s}_i^z(x,t) \hat{s}_0^z(0,0) \rangle_{\delta}^C$, where $\langle \hat{s}_i^z(x,t) \hat{s}_0^z(0,0) \rangle_{\delta}^C = \langle \hat{s}_i^z(x,t) \hat{s}_0^z(0,0) \rangle_{\delta} - \langle \hat{s}_i^z(x,t)\rangle_\delta \langle \hat{s}_0^z(0,0) \rangle_{\delta}$ and the expectation value is taken with respect to the infinite temperature state with average magnetization $\delta/2$~\cite{De_Nardis_2019_2}:
\begin{align}
    \int \mathrm{d}x \:x^2 C(x,t) = \mathcal{D} t^2 + \mathcal{L} t + o(t),
\end{align}
where $\mathcal{D}$ is the Drude weight and the left hand side is the variance of $C(x,t)$ or the mean squared displacement (MSD).
Note that for unmagnetized transport with $\delta = 0$, $\langle \hat{s}_i^z(x,t)\rangle_\delta = \langle \hat{s}_0^z(0,0) \rangle_{\delta} = 0$.
The spin gradient, Eq.~\eqref{supp-eq:gradient}, starting from the magnetization biased domain wall, Eq.~\eqref{supp-eq:charged_spin_DW}, gives the connected spin two-point function, albeit with an improper normalization constant.
Thus to get the properly normalized two-point function and account for the background charge, we normalize $\sum_x C(x,0) = \sum_x C(x,t) = 1/4-\delta^2/4$.
Note that imposing this condition on $\sum_x C(x,t) = \sum_x C(x,0)$ as spin is a conserved quantity and is well conserved (typically $|1 - \sum_x C(x,t)| < 1.e-8$) in our numerics.

\begin{figure}
    \centering
    \includegraphics{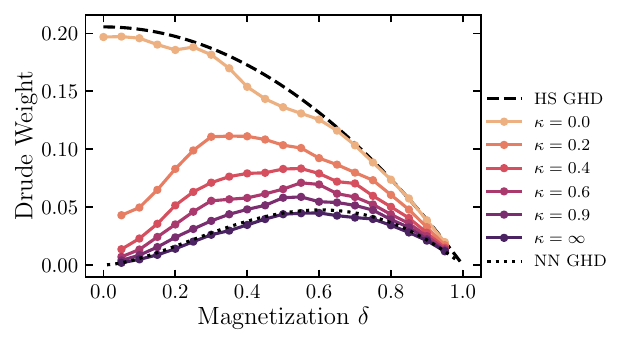}
    \caption{Drude weight for spin transport in a charged background with average magnetization $2 \langle s^z \rangle = \delta$ for Haldane-Shastry, Inozemtsev, and Heisenberg models. Generalized hydrodynamic results are shown where available, showing good agreement across the range of $\delta$.}
    \label{supp-fig:Drude_weight}
\end{figure}

We now study spin transport for various $\delta\in[0,1]$ and $\kappa\in[0, \infty]$ by evolving the weak, charged domain wall and extracting the Drude weight by fitting the MSD; see Fig.~\ref{supp-fig:Drude_weight}.
We compare to GHD results for the nearest-neighbor~\cite{De_Nardis_2020} and Haldane-Shastry~\cite{Bulchandani_2024} model, finding reasonable to good agreement across all $\delta$.
First, one finds a non-zero Drude weight for all models with finite $\delta$, indicating that spin transport is ballistic with a charged background.
Next, we find that the Drude weight for the Inozemtsev models has a maximum at intermediate $\delta$, as transport with $\delta=0$ is KPZ-like superdiffusive and transport with $\delta = 1$ is frozen.
Finally, we note that the $\kappa=0.2$ Inozemtsev model appears to have a finite Drude weight even as $\delta\rightarrow0^+$, but this is due to the slow crossover from ballistic to diffusive transport.
Here we simulate to time $Jt = 200$ with $\chi=128$, so a simulation to later time will likely reduce this numerical artifact.

\bibliography{refs}